\RequirePackage{desc-tex/styles/docswitch}

\documentclass[apj]{emulateapj}

\usepackage{desc-tex/styles/lsstdesc_macros}
\usepackage{physics, amsmath, bm}
\usepackage{graphicx}
\usepackage{savesym}
\usepackage{tikz}
\newcommand*\circled[1]{\tikz[baseline=(char.base)]{
            \node[shape=circle,draw,inner sep=2pt] (char) {#1};}}
\savesymbol{tablenum}
\restoresymbol{SIX}{tablenum}
\usepackage{booktabs, tabularx, multirow, wasysym}
\usepackage{fontawesome, hyperref}
\newcommand{\adriano}[1]{{\color{black}#1}}
\newcommand{\liz}[1]{{\color{black}#1}}
\newcommand{\chris}[1]{{\color{black}#1}}

\usepackage{array}
\newcolumntype{M}{>{\centering\arraybackslash}m{2.cm}}
\newcolumntype{L}{>{\centering\arraybackslash}m{1.5cm}}
\newcolumntype{K}{>{\centering\arraybackslash}m{1.1cm}}
\newcolumntype{Z}{>{\centering\arraybackslash}m{3.5cm}}
\newcolumntype{J}{>{\centering\arraybackslash}m{0.9cm}}
\graphicspath{{./}{./figures/}}

\begin{document}

\title{Hierarchical Inference of the Lensing Convergence from Photometric Catalogs with Bayesian Graph Neural Networks}

\author{{Ji Won~Park} \altaffilmark{1, 2},
{Simon~Birrer}\altaffilmark{1, 2},
{Madison Ueland}\altaffilmark{1},
{Miles Cranmer}\altaffilmark{3},
{Adriano~Agnello}\altaffilmark{4},
{Sebastian~Wagner-Carena}\altaffilmark{1, 2},
{Philip~J.~Marshall}\altaffilmark{1, 2},
{Aaron~Roodman}\altaffilmark{1, 2},
{the~LSST~Dark~Energy~Science~Collaboration}}
\altaffiltext{1}{Kavli Institute for Particle Astrophysics and Cosmology, Department of Physics, Stanford University, Stanford, CA, USA}
\altaffiltext{2}{SLAC National Accelerator Laboratory, Menlo Park, CA, USA}
\altaffiltext{3}{Princeton University, Princeton, NJ, USA}
\altaffiltext{4}{DARK, Niels Bohr Institute, University of Copenhagen, Copenhagen, Denmark}


\begin{abstract}
We present a Bayesian graph neural network (BGNN) that can estimate the weak lensing convergence ($\kappa$) from photometric measurements of galaxies along a given line of sight. 
The method is of particular interest in strong gravitational time delay cosmography (TDC), where characterizing the ``external convergence'' ($\kappa_{\rm ext}$) from the lens environment and line of sight is necessary for precise Hubble constant ($H_0$) inference.
Starting from a large-scale simulation with a $\kappa$ resolution of $\sim$1$'$, we introduce fluctuations on galaxy-galaxy lensing scales of $\sim$1$''$ and extract random sightlines to train our BGNN. 
We then evaluate the model on test sets with varying degrees of overlap with the training distribution. 
For each test set of 1,000 sightlines, the BGNN infers the individual $\kappa$ posteriors, which we combine in a hierarchical Bayesian model to yield constraints on the hyperparameters governing the population. 
For a test field well sampled by the training set, the BGNN recovers the population mean of $\kappa$ precisely and without bias, resulting in a contribution to the $H_0$ error budget well under 1\%. 
In the tails of the training set with sparse samples, the BGNN, which can ingest all available information about each sightline, extracts more $\kappa$ signal compared to a simplified version of the traditional method based on matching galaxy number counts, which is limited by sample variance.
Our hierarchical inference pipeline using BGNNs promises to improve the $\kappa_{\rm ext}$ characterization for precision TDC. 
The implementation of our pipeline is available as a public Python package, \textsc{Node to Joy} \href{https://github.com/jiwoncpark/node-to-joy}{\faicon{github}}.

\end{abstract}


\section{Introduction}
\label{sec:intro}
\chris{Galaxies in the sky may} appear to us distorted in shape and magnified (or de-magnified) in brightness. The responsible phenomenon is \textit{gravitational lensing}: massive structures along our line of sight curve spacetime, thereby gravitationally bending the path of light as it travels to us from the background source. When measured from a large sample of galaxies, the lensing distortions probe the total matter distribution in the Universe, both luminous and dark. A quantity of particular importance in cosmology and galaxy evolution studies is the convergence ($\kappa$), defined as the total integrated mass along a line of sight weighted by the lensing efficiency, which peaks for mass located roughly midway between us and the source. 

Reconstructing a reliable map of the density field $\kappa$ across the sky, or \textit{mass mapping}, can inform the relationship between luminous matter and dark matter, known as the \textit{galaxy-halo connection}. \adriano{Mass maps have been jointly analyzed with maps of stellar mass, galaxies, and galaxy clusters to constrain the galaxy bias---the statistical relation between the distribution of galaxies and matter \citep[e.g.][]{chang2016galaxy}.} See \cite{wechsler2018connection} for a review. 

Mass maps can also improve weak lensing cosmology when used alongside shear maps. Traditional analyses using two-point correlations of the directly measurable shear field are limited to modeling zero-mean Gaussian random fields in the matter density. Extending to higher-order statistics can yield much more information about structure formation, including any non-Gaussianities in the early Universe that would propagate at large scales into the late Universe, as well as those that may arise at small scales later due to nonlinear gravitational collapse \citep{jeffrey2018improving}. 

For reconstructing wide-field mass maps, variants of the direct shear-to-convergence inversion algorithm by \cite{kaiser1993mapping} (``KS'') are commonly used \citep{van2013cfhtlens, vikram2015wide, chang2015wide, liu2015cosmology, chang2018dark, oguri2018two}. The KS method does not account for missing data or noise, \chris{so there have also been efforts to bring in different priors about the convergence field \citep[e.g.][]{wiener1949extrapolation, leonard2014glimpse, lanusse2016high}. A related line of work deals with magnification rather than convergence \citep[e.g.][]{gunnarsson2006corrections,menard2010measuring, morrison2012tomographic}}.

Most sightlines in the Universe are only weakly lensed. \chris{But in the rare event that a halo lines up in front of a source, the lensing is strong enough that the source appears \chris{copied} into multiple images.} The photons from each image arrive at our telescope with relative time delays. If the strongly lensed source is time-variable, e.g. a quasar or a supernova, the time delays can be measured and fed into a cosmographic analysis to infer the Hubble constant ($H_0$) \citep{refsdal1964possibility}. Typically in time delay cosmography, the external lensing perturbations are approximated as a quadrupole lens parameterized by external convergence ($\kappa_{\rm ext}$) and shear. The external shear can be constrained by modeling the lenses from \chris{imaging data}, but $\kappa_{\rm ext}$ cannot---a manifestation of the \textit{mass-sheet degeneracy} \citep{falco1985model}. We therefore require external tracers of the density field in addition to the strong lens modeling.

\cite{suyu2010dissecting} \chris{and \cite{fassnacht2011galaxy}} pioneered the method of estimating $\kappa_{\rm ext}$ based on galaxy number counts. In brief, the method involves (1) counting the galaxies around a lens system, (2) comparing the resulting counts against those of a control field, \adriano{e.g.\ the Cosmic Evolution Survey (COSMOS) field \citep{scoville2007cosmos} and the Canada-France Hawaii Telescope Lensing (CFHTLenS) Survey \citep{cfhtlens}}, to obtain relative counts, and (3) selecting lines of sight of similar relative counts, along with their associated convergence values, from a numerical N-body simulation with an assumed galaxy-halo connection. The matched $\kappa$ samples can then be interpreted as $\kappa_{\rm ext}$ posterior samples for a sightline, where the data for the sightline has been reduced to number counts as a summary statistic. \adriano{This method of matching summary statistics rather than evaluating a specific likelihood function can be cast as approximate Bayesian computation (ABC) \citep{birrer2019h0licow}. Note that the galaxy-halo connection encoded in the the N-body simulation and the semi-analytic galaxy evolution model is a key source of uncertainty in the $\kappa$ constraints.}  

Since then, the method has evolved to accommodate summary statistics beyond simple galaxy counts. \cite{suyu2013two} used the external shear inferred from lens modeling as an additional constraint. \cite{greene2013improving} explored schemes to weight number counts by the galaxy redshift, stellar mass, and projected separation from the line of sight. The additional information improved the $\kappa_{\rm ext}$ constraints for the most overdense lines of sight but helped little for the less dense ones. They found that the underdense sightlines yielded the best $\kappa_{\rm ext}$ constraints, at a residual uncertainty of $\sigma_{\kappa_{\rm ext}} \lesssim 0.03$, corresponding to an uncertainty on $H_0$ comparable to that from lens modeling. Alternatively, \cite{collett2013reconstructing} reconstructed $\kappa_{\rm ext}$ using a galaxy halo model, calibrating for the effect of dark structures and voids. Similarly to \cite{greene2013improving}, they also reported the most precise estimates for environments with low $\kappa_{\rm ext}$. \cite{mccully2014new} devised the ``flexion shift'' metric quantifying the contribution of a line-of-sight object to the lens potential. For the ground-truth $\kappa_{\rm ext}$, all of these studies used the 1.7 deg$^2$ mock sky maps of $\kappa_{\rm ext}$ obtained by raytracing through the Millennium Simulation \citep{springel2005simulations, hilbert2009ray}. \chris{The analysis methods shed light on the spectroscopic data previously collected around strong lenses for line-of-sight assessment \citep[e.g.][]{Fassnacht:2002df,fassnacht2006mass,momcheva2006spectroscopic}}. 

To date, seven lensed quasars have undergone cosmographic analysis following a fairly standard procedure. The $\kappa_{\rm ext}$ constraints were generated on an individual lens level by matching relative number counts, both with and without weighting by the inverse projected separation ($1/r$) \citep{greene2013improving, rusu2017h0licow, rusu2020h0licow, chen2019sharp, buckley2020strides}. The H0LiCOW and SHARP collaborations measured $H_0$ from six lenses with \chris{the median} $\kappa_{\rm ext}$ ranging from $-0.006$ (PG 1115+080) to $0.1$ (B1608+656) \citep{suyu2010dissecting, suyu2014cosmology, wong2017h0licow, birrer2019h0licow, rusu2020h0licow, chen2019sharp, wong2019h0licow}. The uncertainty contribution of $\kappa_{\rm ext}$ to the individual $H_0$ error budget ranged from ${\sim}2.7$\% to ${\sim}6.5$\%. The STRIDES collaboration measured $H_0$ to 3.9\% precision from one additional lens inside a particularly underdense environment, of estimated $\kappa_{\rm ext} \sim -0.04$ \citep{shajib2020strides}. 

The H0LiCOW/COSMOGRAIL/STRIDES/SHARP collaborations (hereafter TDCOSMO\footnote{\url{www.tdcosmo.org}} collaboration) currently constrain $\kappa_{\rm ext}$ for each lens independently when inferring $H_0$ from the combined lens sample. With this standard procedure, TDCOSMO achieved ${\sim}$2\% precision on $H_0$ from the seven lenses \citep{rusu2020h0licow, wong2019h0licow, chen2019sharp, shajib2019every, millon2020tdcosmo}. The uncertainty increased to ${\sim}5$\% when the lens mass models were made significantly more flexible \citep{birrer2020tdcosmohierarchical}. \chris{In relaxing assumptions on the lens mass model, \cite{birrer2020tdcosmohierarchical} marginalized over the population offset in the mass sheet that is \textit{internal} to the main deflector, which manifests as a transform of its mass profile. While these analyses did not hierarchically account for the \textit{external} aspect of the mass sheet due to line-of-sight structure, the interpretation of the mass sheet can be extended to include external effects.}

Hierarchically inferring the population mean of $\kappa_{\rm ext}$ and marginalizing over it is important, because any bias in the assumed population mean of $\kappa_{\rm ext}$ directly biases $H_0$. Note that the mean $\kappa_{\rm ext}$ may not necessarily vanish for an ensemble of lenses due to selection effects, e.g. lens galaxies tend to lie in groups \citep{blandford2000modeling}, causing a slight preference for lens systems with overdense lines of sight \chris{\citep{fassnacht2011galaxy,collett2016observational}}. \chris{Any discrepancy between our prior and the actual $\kappa_{\rm ext}$ population can result in bias, so we must hierarchically infer the actual population statistics and account for it in TDC analysis.} The Legacy Survey of Space and time (LSST) at the Vera Rubin Observatory \chris{is predicted to discover} tens of thousands of lenses \citep{collett2016observational}, among them ${\sim}$8,000 lensed quasars \citep{oguri2010gravitationally}. \chris{As we scale up to such a large sample, hierarchical inference of $\kappa_{\rm ext}$ can combine the information from all sightlines, each with limited signal on $\kappa_{\rm ext}$, to minimize systematic bias on $H_0$.}

Although uncertainties in other components of time delay cosmography, such as lens modeling and time delay measurements, have reduced over time, $\kappa_{\rm ext}$ remains a significant source of uncertainty on $H_0$. The primary drawback of matching relative number counts to estimate $\kappa_{\rm ext}$ is that the wealth of photometric observations for each galaxy---such as the measured magnitudes in each bandpass filter, positions, and shapes---become reduced to a few numbers for the whole sightline. We seek to extract information more efficiently by enlisting a neural network to process all available observations. Namely, we design a Bayesian graph neural network (BGNN) that can take in the measurements of a variable number of galaxies observed around a given line of sight and output the full posterior probability density function (PDF) over $\kappa$. Training the BGNN on a galaxy catalog with associated $\kappa$ labels allows it to implicitly learn the distribution of dark matter in galaxies and clusters \adriano{encoded in the underlying simulation.} This precludes the need to manually correct for the mass that is not in galaxy or cluster halos or assume overly simple models for the lensing mass, \chris{as done in \cite{collett2013reconstructing}.}

This paper seeks to improve the $\kappa_{\rm ext}$ estimation on three fronts: (1) proposing the BGNN as a novel $\kappa_{\rm ext}$ estimation engine, (2) combining the BGNN constraints into hierarchical inference of the population $\kappa_{\rm ext}$, and (3) enhancing the lensing scales of large-scale simulations to adequately describe lensing effects on the scale of galaxy-scale strong lenses. We ask the following questions:
\begin{itemize}
    \item Is the Bayesian graph neural network (BGNN) capable of accurately and precisely estimating $\kappa$ for individual sightlines?
    \item When propagated into a hierarchical inference framework, do the BGNN-inferred posteriors lend themselves to precise and accurate recovery of hyperparameters governing $\kappa$ in the population? This investigation seeks to quantify population-level selection effects in regimes where insufficient information is available on an individual sightline-by-sightline basis.
    \item How does the BGNN compare with the ABC technique of matching number counts?
    \item Which environments lend themselves to better $\kappa$ constraints? At what extremes does this method break?
\end{itemize}

To answer these questions, we train our BGNN on a realistic galaxy catalog with associated $\kappa$ labels and design a suite of numerical experiments to validate our approach. 

We organize this paper as follows. Section \ref{sec:cosmography} provides an introduction to time delay cosmography, with a focus on the impact of $\kappa_{\rm ext}$ on $H_0$ inference. Section \ref{sec:methods} details our hierarchical inference pipeline. Section \ref{sec:results} reports our $\kappa$ recovery results on 1,000 test sightlines. Section \ref{sec:conclusion} situates our findings within the context of time delay cosmography and suggests next steps.

We reserve a bulk of the details about our simulations to the Appendix (Section \ref{sec:lensing_formalism}). There, we start with a primer on the lensing formalism and describe how we use multi-plane raytracing to post-process an existing galaxy catalog and $\kappa$ map, produced in the weak lensing resolution, for strong lensing applications. 

\chris{Throughout, we will present comparisons with the method of matching summary statistics, in order to highlight the difference between a neural net and an ABC-based algorithm. Our implementation of the summary statistics matching does not represent the TDCOSMO implementation. TDCOSMO uses orders of magnitude more sightlines, allowing them to match simultaneously to two sets of summary statistics, whereas we match to one summary statistic at a time. Additionally, TDCOSMO incorporates external shear information from the lens modeling to further constrain their $\kappa_{\rm ext}$. This is not possible within our context, because we only operate on sightlines without strong lensing.}

We release with this publication the open-source Dark Energy Science Collaboration
(DESC) Python package \textsc{Node to Joy}\footnote{\faicon{github} \url{http://github.com/jiwoncpark/node-to-joy}}. This package implements the complete pipeline, including
the structure-enhanced raytracing, training and evaluation of the BGNN, comparison with summary statistics matching, and the hierarchical inference analysis.

\section{Strong Lensing Time Delay Cosmography}
\label{sec:cosmography}
Let us begin by reviewing the basic principles of time delay cosmography \citep{refsdal1964possibility}. Readers are referred to recent reviews, e.g. \cite{treu2016time}, for more details. When light rays from a distant source are deflected by some foreground lens, the travel time from the source to the observer depends on both their path lengths and the gravitational potential they traverse. Assuming a single thin lens, the excess time delay of an image at position $\boldsymbol{\theta}$ originating from a source at position $\boldsymbol{\beta}$ relative to an unperturbed path is
\begin{align} \label{eq:excess_time_delays}
    t(\boldsymbol{\theta}, \boldsymbol{\beta}) = \frac{D_{\Delta t}}{c} \phi(\boldsymbol{\theta}, \boldsymbol{\beta}).
\end{align}
Here,
\begin{align} \label{eq:fermat_potential}
\phi(\boldsymbol{\theta}, \boldsymbol{\beta}) = \left[\frac{ (\boldsymbol{\theta} -\boldsymbol{\beta})^2}{2} - \psi(\boldsymbol{\theta})\right]
\end{align}
is the Fermat potential \citep{schneider1985new, blandford1986fermat}, defined for the lensing potential $\psi(\boldsymbol{\theta})$, a two-dimensional projection of the three-dimensional Newtonian potential $\Phi(\eta \bm{\theta}, \eta)$: 
\begin{align} \label{eq:lensing_potential}
    \psi(\bm{\theta}) = 2 \int_{0}^{\eta} d\eta' \frac{D(\eta - \eta')}{D(\eta') D(\eta)} \Psi(\eta' \bm{\theta}, \eta').
\end{align}
$D_{\Delta t}$ is the time delay distance:
\begin{align} \label{eq:time_delay_distance}
    D_{\Delta t} = (1 + z_d) \frac{D_{\rm d} D_{\rm s}}{D_{\rm ds}}
\end{align}
where $z_d$ is the redshift of the deflector galaxy and $D_{\rm d}$, $D_{\rm s}$, $D_{\rm ds}$ are the angular diameter distances to the deflector, to the source, and between the deflector and the source, respectively. We have that $D_{\Delta t} \propto 1/H_0$.

Additionally, structures along the line of sight (LOS) to the strong lens introduce additional weak lensing perturbations. The resulting convergence affects the time delays while keeping imaging observables the same under linear transformations of the lens equation---the so-called mass-sheet degeneracy (MSD).
As shown by \cite{falco1985model}, remapping the reference mass distribution $\kappa$, for any scalar $\lambda$, as
\begin{equation}
    \label{eq:msd}
    \kappa_{\lambda}(\theta) = \lambda \kappa(\theta) + (1 - \lambda)
\end{equation}
 while isotropically scaling the source plane as $\beta \rightarrow \lambda \beta$ results in the same dimensionless observables (image pixel values) but different time delays. The infinite family of solutions to the lens equation results in a range of inferred properties of the deflector and the source.
 
 The mass sheet parameter $\lambda$ in Equation \ref{eq:msd} may be internal to the main deflector, affecting its kinematics, or it may stem from the external line-of-sight structure. It is common practice to express the external portion $\lambda_{\rm ext}$ as an equivalent additional mass sheet at the redshift of the main deflector with uniform surface mass density, called external convergence and denoted $\kappa_{\rm ext}$ \citep{schneider1997cosmological, keeton1997shear}. In terms of $\lambda_{\rm ext}$, we have $\kappa_{\rm ext} = 1 - \lambda_{\rm ext}$. 

If we were to use a model not accounting for $\kappa_{\rm ext}$ (i.e. fixing $\kappa_{\rm ext}$=0) to infer the time delay distance $D_{\Delta t}^{(0)}$, then the true time-delay distance can be recovered using a separately constrained $\kappa_{\rm ext}$ as:
\begin{align} \label{eq:kappa_D_dt_relation}
    D_{\Delta t} = \frac{D_{\Delta t}^{(0)}}{1 - \kappa_{\rm ext}}
\end{align}
In terms of $H_0$, this relation is:
\begin{align} \label{eq:kappa_H0_relation}
    H_0 = (1 - \kappa_{\rm ext}) H_0^{(0)}.
\end{align}
Underestimating $\kappa_{\rm ext}$ thus leads to an upward bias on $H_0$ and vice versa.

The external convergence $\kappa_{\rm ext}$ is a product of three different convergence values \citep{birrer2020tdcosmohierarchical, 2021fleury}:
\begin{align}
    1 - \kappa_{\rm ext} = \frac{(1-\kappa_{\rm d})(1-\kappa_{\rm s})}{1 - \kappa_{\rm ds}},
\end{align}
where $\kappa_{\rm s}$, $\kappa_{\rm d}$, $\kappa_{\rm ds}$ correspond to the integrated convergence along the strong lens line of sight from the observer to the source, from the observer to the strong lens, and from the strong lens to the source, respectively.  
These component convergence values transform the background angular diameter distances ($D^{\rm BG}$) from the homogeneous background metric given by the cosmological model, without any perturbations, into the angular diameter distances along the specific line of sight of the strong lens ($D^{\rm SL}$). That is,
\begin{align}
    D^{\rm SL}_{\rm s/d/ds} = (1 - \kappa)D^{\rm BG}_{\rm s/d/ds}.
\end{align}
We apply our method only to $\kappa_{\rm s}$, but it can be generalized to accommodate all three angular diameter distances.

Because each sightline contains limited information about $\kappa_{\rm ext}$, the inference of $\kappa_{\rm ext}$ necessitates a hierarchical approach. For a sample of $N_{\rm lens}$ lenses, we can apply the simplified and analytical error propagation in \cite{birrer2021hubble} to evaluate the contribution of $\kappa_{\rm ext}$ to the overall $H_0$ error budget. We decompose the $H_0$ uncertainty into two terms: one term capturing global shifts in the inferred $H_0$ due to the mass sheet $\lambda$ and another term incorporating uncertainties introduced by individual lenses, including uncertainties from the time delay measurements and the inferred Fermat potentials of main deflectors. The fractional uncertainty on $H_0$ can be written as
\begin{align} \label{eq:frac_uncertainty_H0_breakdown}
    \left( \frac{\sigma(H_0)}{H_0} \right)^2 \approx \underbrace{\left(\frac{\sigma(\bar \lambda)}{\bar \lambda} \right)^2}_{\text{\circled{1} : global mass sheet}} + 
    \underbrace{\left(\frac{\sigma(H_0^{(0)})}{H_0^{(0)}} \right)^2}_{\text{\circled{2}: indiv. err.}},
\end{align}
where $\bar \lambda$ refers to the population mean in $\lambda$ and $\sigma(x)$ to the 1$\sigma$ uncertainty on \chris{the quantity $x$}.  

Note that $\bar \lambda $  of Equation \ref{eq:frac_uncertainty_H0_breakdown} includes both the internal and external mass sheets. Other probes that are sensitive to both can also constrain $\bar \lambda$ but would not allow such a separation in the physical interpretation. The contribution of the external portion $\bar \lambda_{\rm ext}$ to term \circled{1} of Equation \ref{eq:frac_uncertainty_H0_breakdown} is
\begin{align} \label{eq:mean_contrib_H0}
    \left (\frac{1 - \sigma(\bar \lambda_{\rm ext})}{\bar \lambda_{\rm ext}} \right)^2 = \left( \frac{\sigma(\bar \kappa_{\rm ext})}{1 - \bar \kappa_{\rm ext}}\right)^2.
\end{align}
This represents the main contribution of $\kappa_{\rm ext}$ to the overall $H_0$ error budget. There is a smaller, second-order contribution that enters term \circled{2} in Equation \ref{eq:frac_uncertainty_H0_breakdown}: 
\begin{align} \label{eq:scatter_contrib_H0}
     \frac{1}{N_{\rm lens}} \left( \frac{\sigma_{\kappa_{\rm ext}}}{1 - \bar \kappa_{\rm ext}}\right)^2,
\end{align}
where $\sigma_{\kappa_{\rm ext}}$ denotes the $1\sigma$ scatter in the population $\kappa_{\rm ext}$. This contribution scales inversely with $N_{\rm lens}$ and becomes subdominant to other sources of uncertainty, especially in the large-$N_{\rm lens}$ setting. 

Suppose we estimate $\bar \kappa_{\rm ext}$ for the population mean in $\kappa_{\rm ext}$ but the truth lies offset from this, at $\bar \kappa_{\rm ext, true}$. It follows from Equation \ref{eq:kappa_H0_relation} that the fractional bias on $H_0$ introduced by an erroneous estimate $\bar \kappa_{\rm ext}$ is
\begin{align} \label{eq:h0_bias}
    \textrm{Frac. bias on $H_0$} = \frac{\bar \kappa_{\rm ext, true} - \bar \kappa_{\rm ext}}{1 - \bar \kappa_{\rm ext, true}}.
\end{align}
That is, the error on $\bar \kappa_{\rm ext}$ biases $H_0$ almost directly.

In this paper, we perform hierarchical inference to obtain the full posterior PDF over the population $\kappa_{\rm ext}$ statistics, $\bar \kappa_{\rm ext}$ and $\sigma_{\kappa_{\rm ext}}$ (see Section \ref{sec:hierarchical_inference}). Equation \ref{eq:mean_contrib_H0} and, to a lesser extent, Equation \ref{eq:scatter_contrib_H0} allow us to contextualize our hierarchical constraints within the overall $H_0$ error budget.

\section{Methods} \label{sec:methods}

\begin{figure*}[!htb]
\includegraphics[width=1\textwidth,trim={0 1.2cm 0 0},clip]{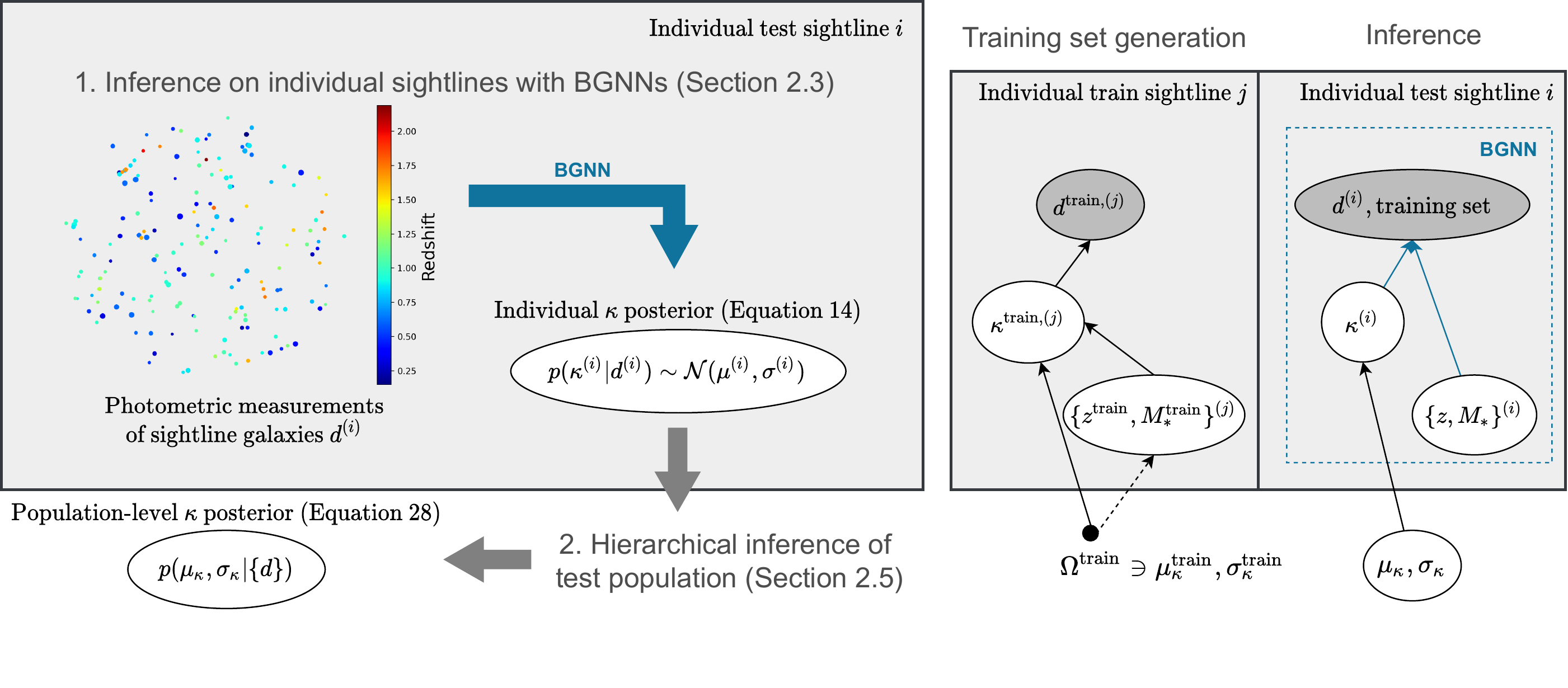}
\caption{Left: Our BGNN takes as input the photometric measurements, $d^{(i)}$, of galaxies comprising a test sightline $i$ and outputs the posterior on $\kappa$, $p(\kappa^{(i)}|d^{(i)})$. After repeating this for all the test sightlines, we combine the posteriors into a hierarchical Bayesian model to constrain the $\mu_{\rm test}, \sigma_{\rm test}$ governing the test population. Right: simplified dependence relations shown as PGMs. Dots denote fixed values, shaded ovals observations, and unshaded ovals random variables. During training set generation, the set of assumptions governing the training set, $\Omega_{\rm train}$, are used to realize specific values of $\{z, M_\star\}$ and $\kappa$ for individual sightlines. The photometric observations of the field of view for that sightline, $d$, depend on $\kappa$, among other quantities not shown. During inference, the trained BGNN infers $\kappa$ from $d$. The resulting posteriors for all test sightlines are used to recover $\mu_{\rm test}, \sigma_{\rm test}$ of the test population.}
\label{fig:pipeline_diagram}
\end{figure*}

In this section, we outline the methodology for the various stages of our $\kappa$ inference pipeline. Figure \ref{fig:pipeline_diagram} illustrates our pipeline as a flowchart and probabilistic graphical models (PGMs) for training and inference. In Section \ref{sec:simulated_data}, we describe the simulated training data. In particular, we state the numerical choices made when postprocessing the existing galaxy catalog and a map of weak lensing $\kappa$ to generate the training set suitable for strong lensing applications. Then, in Section \ref{sec:bgnn}, we explain how the BGNN extracts the $\kappa$ posteriors from individual sightlines. Alternatively, the traditional summary statistics method can be used to obtain the posteriors, as described in Section \ref{sec:summary_stats}. Lastly, the $\kappa$ posteriors, whether from the BGNN or the summary statistics matching, enter the hierarchical inference framework to yield constraints on hyperparameters governing the population; Section \ref{sec:hierarchical_inference} describes this process. 

\subsection{Simulated data} \label{sec:simulated_data}
Our training set consists of a galaxy catalog queried around random sightlines and the associated $\kappa$ labels. Both the input catalog and target labels are derived from \textsc{CosmoDC2}, a catalog based on a \adriano{cosmological N-body simulation and a semi-analytic galaxy evolution model \citep{korytov2019cosmodc2}.} The catalog contains the $\kappa_{\rm WL}$ computed in the weak lensing (WL) regime ($\delta \sim 1'$). We perform additional raytracing on top of the \textsc{CosmoDC2} halos to introduce fluctuations on a finer angular ($\delta \sim 1''$) resolution required for strong lensing studies. Section \ref{sec:structure_enhanced} details this structure-enhanced raytracing procedure. The surface mass densities were estimated by rendering halos with \adriano{the Navarro-Frenk-White (NFW) mass profile \citep{navarro1997universal}}, described in Section \ref{sec:halo_rendering}. Section \ref{sec:galaxy_catalog} briefly summarizes the semi-analytic models used to generate the galaxy catalogs. We train our BGNN on the photometric information associated with \textsc{CosmoDC2} sightlines and the $\kappa$ values that we postprocessed to the strong lensing resolution. \chris{The source redshift was fixed at $z_{\rm src} = 2$, typical of lenses expected to be discoverable in upcoming wide-area surveys \citep{collett2015population}. This was a choice also made in \citep{li2020impact}.}

\adriano{The goal of this paper is to validate our hierarchical $\kappa$ inference pipeline within a toy testbed, by explicitly treating the training distribution as a prior and conducting retrieval tests using carefully chosen test sets.} We choose a broad Gaussian distribution for the training $\kappa$ to enable an analytical treatment of the BGNN prior. \chris{As we will see in Section \ref{sec:hierarchical_inference}, Gaussian priors are especially convenient in our hierarchical inference context, because it has support everywhere in $\mathbb{R}$.} We subsampled 200,000 sightlines from the phenomenological set of 600,000 \textsc{CosmoDC2} sightlines to follow $\kappa \sim \mathcal{N}(0.01, 0.04)$ where $\mathcal{N}(\mu, \sigma)$ is a normal distribution with mean $\mu$ and standard deviation $\sigma$. The validation set contains 1,000 examples, drawn from the same distribution as the training set. 

Our primary target label for each sightline is $\kappa$. We additionally design per-galaxy labels, the redshift ($z$) and stellar mass ($M_\star$) of each galaxy included in the \textsc{CosmoDC2} catalog. We are motivated by the idea that training the BGNN to simultaneously predict $z, M_\star$ of each galaxy as well as $\kappa$ may improve the inductive bias of the network for $\kappa$ inference. We anticipate that it is important for the network to understand galaxy redshifts well, so that it can predict the relative $\kappa$ contribution of the associated halos to the overall $\kappa$ for the sightline, i.e. the lensing efficiency. The network must also internally learn to exclude a galaxy/halo from consideration if it is behind the source redshfit of $z_{\rm src}=2$. We included stellar masses as a per-galaxy label to serve as proxies for halo masses, which were only available for central galaxies (populating main halos) in the \textsc{CosmoDC2} catalog and not for satellite galaxies (populating subhalos). 

\begin{figure*}[!htb]
\includegraphics[width=1\textwidth]{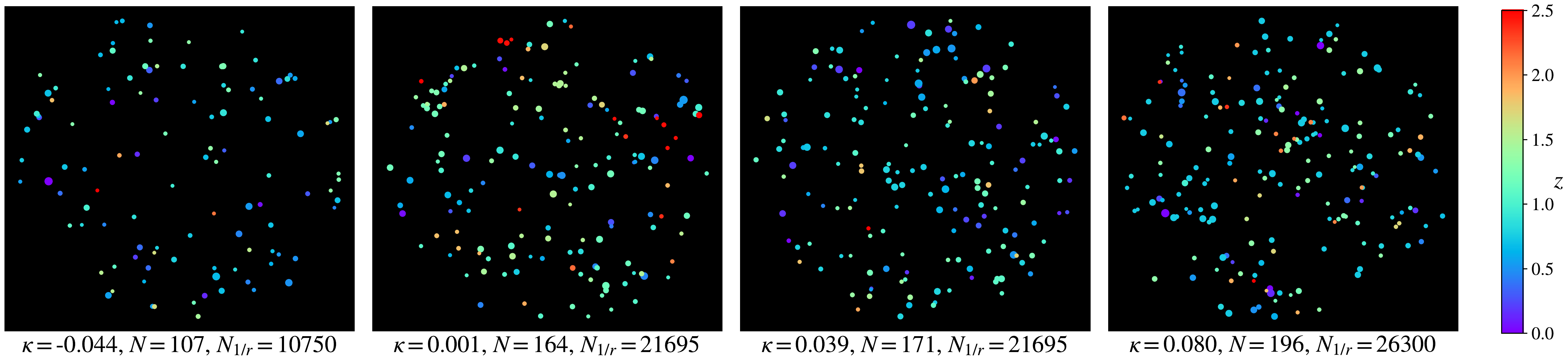}
\caption{Illustration of $\kappa$ in the training sightlines. Each dot is a galaxy meeting the photometric cut $i < 25.3$. Colors of dots indicate redshift $z$ and sizes of dots increase linearly with decreasing $i$-band magnitude. From left to right, the true $\kappa$ increases at approximately the centers of test sets C1 through C4 (-0.04, 0, 0.04, 0.08). The unweighted number counts $N_{\rm sim}$ and distance-weighted number counts $N_{1/r, {\rm sim}}$ (Equation \ref{eq:n_inv_dist_def}) also generally increase with increasing $\kappa$. The source was located at $z=2$ for all the sightlines. Best viewed in color.}
\label{fig:gallery}
\end{figure*}

Our input is the photometric catalog of galaxies queried within an aperture of radius 1$'$ around a sightline. The input features of each galaxy were derived from the \textsc{CosmoDC2} sky position and magnitudes in \textit{ugrizY}. We compute the relative sky position of each galaxy from the central line of sight located at ${\rm RA}_{\rm LOS}$, ${\rm dec}_{\rm LOS}$ and transform the distances to the flat sky. 

Gaussian photometric errors corresponding to 5-year LSST conditions were simulated and added to the true \textit{ugrizY} magnitudes. We adopted the analytic magnitude-dependent model for the 1-$\sigma$ uncertainties described in \cite{ivezic2019lsst}. The model includes dependencies on true magnitudes as well as band-dependent parameters like sky brightness, seeing, atmospheric extinction, instrumental noise, and the number of visits. Figure \ref{fig:gal_err} shows that our error model agrees with the estimated photometric uncertainties in the DC2 catalog \citep{abolfathi2021lsst}, which simulates 5 years of the planned 10-year LSST survey. In addition, we applied a detection cut of $i<25.3$, corresponding to the 10-year $i$-band gold sample \citep{gorecki2014new}. 

\adriano{In summary, the input to our model was a variable-sized set of galaxies located within 1$'$ of a sightline. Each galaxy carried eight features: the relative positional offset from the central line of sight in RA and dec and the \textit{ugrizY} magnitudes with photometric noise added.}

\subsection{Bayesian graph neural network}
\adriano{Bayesian neural networks (BNNs) are probabilistic extensions to standard neural networks \citep{denker1991transforming}. As outlined in \cite{jospin2020hands}, designing a BNN entails choosing a \textit{stochastic model}, which defines the prior on the network weights and the predictive distribution on the target quantities, and a \textit{functional model}, which defines the network architecture. We describe the stochastic model in Section \ref{sec:posterior_inference} and the functional model in Section \ref{sec:network_architecture}.}

\label{sec:bgnn}
\subsubsection{Posterior inference with Bayesian neural networks} \label{sec:posterior_inference}
It is instructive to decompose the uncertainty estimated by BNNs into two types: aleatoric (statistical, \liz{irreducible}) and epistemic (systematic, \liz{reducible}).

Originating from the intrinsic randomness in the data-generating process, the aleatoric uncertainty persists in the limit of infinite training data. This type of uncertainty is explicitly modeled as the width of the distribution over the target quantities. Concretely, let a sightline with photometric observations $\bm{d}$ be given. Suppose it contains $L$ galaxies, indexed $l=1, \dots, L$. Our targets are $\kappa$ and the set of redshifts and stellar masses of the individual galaxies, $\{z, M_\star \} \equiv \{z_l, M_{\star, l} \}_{l=1}^L$. The aleatoric portion of our posterior is thus $p\left(\kappa, \{z, M_\star \}|d, W\right)$, for a set of network weights, $W$.

In this paper, \adriano{we adopt an independently Gaussian predictive distribution for} $\kappa$ and $\{z, M_\star \}$, i.e.
\begin{align} \label{eq:aleatoric_posterior}
     \log p & \left(\kappa, \{z, M_\star \}|\bm{d}, W \right) \nonumber \\
     \propto &  \log \mathcal{N}\left(\kappa |m(\bm{d}, W), s(\bm{d}, W)\right) \nonumber \\ 
     &+ \sum_l \log \mathcal{N} \left(z_l|t_l(\bm{d}, W), u_l(\bm{d}, W) \right) \nonumber \\ 
     &+ \sum_l \log \mathcal{N} \left( M_{\star, l}|v_l(\bm{d}, W), w_l(\bm{d}, W) \right),
\end{align}
where $\mathcal{N}(\cdot|\mu, \sigma)$ denotes the PDF of a Gaussian with mean and standard deviation $\mu, \sigma \in \mathbb{R}$. The BNN predicted the $m, \log s$ and $\{t_l, \log u_l, v_l, \log w_l\}_{l=1}^L$ for each sightline, where $l$ indexes the galaxies in a sightline.
We chose the Gaussian as a first-order approximation to the true posterior to focus on simple $\kappa$ recovery tests. To control for effects from the prior, our training set has been subsampled from the original simulation to follow a Gaussian distribution in $\kappa$ (Section \ref{sec:simulated_data}). \adriano{We train the network to jointly infer the per-galaxy properties $\{z, M_\star\}$ along with $\kappa$ in order to improve the inductive bias of the BGNN \citep{battaglia2018relational}. The idea is that explicitly training the network on $\{z, M_\star\}$ labels will guide its understanding of the per-galaxy contribution to the overall $\kappa$ for the sightline.}


Epistemic uncertainty stems from our limited knowledge of the perfect model. It is reducible in the sense that it can be explained away with sufficient training data. This type of uncertainty is often described as a distribution over the network weights $W$. Each realization of the weights corresponds to an alternative model, so integrating over this learned weight posterior amounts to Bayesian model averaging. Folding in the epistemic uncertainty, we have the full predictive distribution for $\kappa$ of a given sightline: 
\begin{align} \label{eq:predictive_distribution}
    p(\kappa | \bm{d}, \Omega_{\rm train}) = \int p(\kappa| \bm{d} , W) p(W| \Omega_{\rm train}) \ dW
\end{align}
where we have made explicit the dependence on the training set by conditioning on the hyperparameters governing the training prior, $\Omega_{\rm train}$. Not modeling the epistemic uncertainty at all reduces to simple density estimation, where the weight posterior $p(W | \Omega_{\rm train})$ is a delta function. In standard neural networks, which only give point estimates for the target parameters, both $p(\kappa | \bm{d} , W)$ and $p(W |\Omega_{\rm train})$ are delta functions, so the predictive distribution in Equation \ref{eq:predictive_distribution} collapses to a delta function. 

Exact evaluation of the integral in Equation \ref{eq:predictive_distribution} requires averaging over all the weight configurations allowed by $p(W | \Omega_{\rm train})$, which is intractable. There exist \adriano{several approximations} (see, e.g. \citealt{charnock2020bayesian} for a review). Among them, we opt for Monte Carlo (MC) dropout, because it precludes the need to train an ensemble of BNNs \citep{gal2016dropout, kendall2017uncertainties}. MC dropout is a Bayesian interpretation of regular dropout, whereby nodes are randomly dropped out, or set to zero, \adriano{with some tuned probability}. Mathematically, it replaces the true weight posterior $p(W |\Omega_{\rm train})$ with a variational distribution $q_\theta(\hat W |\Omega_{\rm train})$ parameterized by $\theta$ \citep{gal2016dropout}:
\begin{align}
& q_\theta(\hat W |\Omega_{\rm train}) = \prod_i q_\theta(\hat W_i |\Omega_{\rm train}) \nonumber \\
& \hat W_i = W_i \cdot {\rm diag}\left(z_{i, j} \right)_{j=1}^{J_i} \nonumber \\
&z_{i, j} \sim {\rm Bernoulli}(p_i) \nonumber \\
&\theta \equiv \{W_i, p_i \}_{i=1}^L
\end{align}
where $i$ indexes layers of the network and $j$ the nodes in a given layer. Here, $J_i$ denotes the number of nodes in layer $i$, such that the weight matrix for layer $i$ is $W_i \in \mathbb{R}^{J_i \times J_{i-1}}$. When $z_{i,j} = 0$, the input node $j$ in layer $i$ is dropped out. \adriano{In practice, the per-layer dropout probability $p_i$ is replaced with a global dropout probability $p_{\rm drop}$.}

The full predictive posterior with the variational approximation is thus
\begin{align} \label{eq:predictive_distribution_variational}
    p(\kappa | \bm{d},  \Omega_{\rm train}) &= \int p(\kappa| \bm{d}, \hat W) q_\theta(\hat W| \Omega_{\rm train}) \ \bm{d}\hat W.
\end{align}

\subsubsection{Optimization}

The loss function is the negative log evidence lower bound (ELBO) over the $N$=200,000 examples in the training set, $\{\bm{d}^{(n)}, \kappa^{(n)}, \{z, M_\star \}^{(n)}\}_{n=1}^{N}$:
\small
\begin{align} \label{eq:bnn_loss}
    \mathcal{L}(W) = & \sum_{n=1}^{N} \int \log p(\kappa^{(n)}, \{z, M_\star \}^{(n)}|\bm{d}^{(n)},\hat W) q_\theta(\hat W |\Omega_{\rm train}) \bm{d}\hat W \nonumber \\
    & + \text{KL}(q_\theta(\hat W |\Omega_{\rm train})||p(\hat W)) 
\end{align}
\normalsize
where $p(W)$ is a prior on the network weights. To evaluate the first term in an unbiased way, we take a single MC sample $\hat W \sim q_\theta(\hat W |\Omega_{\rm train})$. Then $W$ can be updated via gradient descent with respect to the realized sample. The second KL term is the ``regularization'' term that prevents the weights from deviating too far from our prior. This is intractable in its exact form, but reduces to $L_2$ regularization
\begin{align} \label{eq:reg_term}
    \text{KL}(q_\theta(\hat W_i | \Omega_{\rm train})||p(\hat W_i)) \propto \frac{h^2 (1-p_i)}{2N}||\hat W_i||^2
\end{align}
when we assume a weight prior that can be factorized into a product of Gaussian priors in each layer. The length scale $h$ is a hyperparameter that determines the width of the prior. Note that the dropout probability is also a hyperparameter in the formulation introduced here. It is not optimized along with $W$ during training and was tuned manually as part of the hyperparameter search. We assume the same dropout probability $p_i=p_{\rm drop}$ for every layer $i$. For a given choice of $p_{\rm drop}$, $h$ can be folded into the hyperparameter $\lambda = h^2 (1-p)/(2N)$ controlling the $L_2$ regularization strength. \adriano{We tune $p_{\rm drop}$, the initial learning rate, batch size, $\lambda$, and the number of BGNN layers (network depth) on a sparse random grid as part of our hyperparameter search.}

We train our BNN via minibatch gradient descent using the \textsc{ADAM} optimizer with a weight decay of $\lambda=$1e-4 and batch size of 1,000. The learning rate, initially 1e-3, is halved whenever the validation loss fails to decrease for 5 consecutive epochs. We stop training when the validation loss does not increase for 10 consecutive epochs.

It is common practice to transform the training input and labels so that they fall into a predefined range. This preprocessing step has the effect of facilitating optimization, as it promotes the numerical stability of the network's hidden units and their gradients. The input $\bm{d}$ and target $\kappa$ labels are normalized so that they have a mean of 0 and standard deviation of 1 across the entire training set.

\subsubsection{Network architecture} \label{sec:network_architecture}

\begin{figure*}[!htb]
\includegraphics[width=1\textwidth]{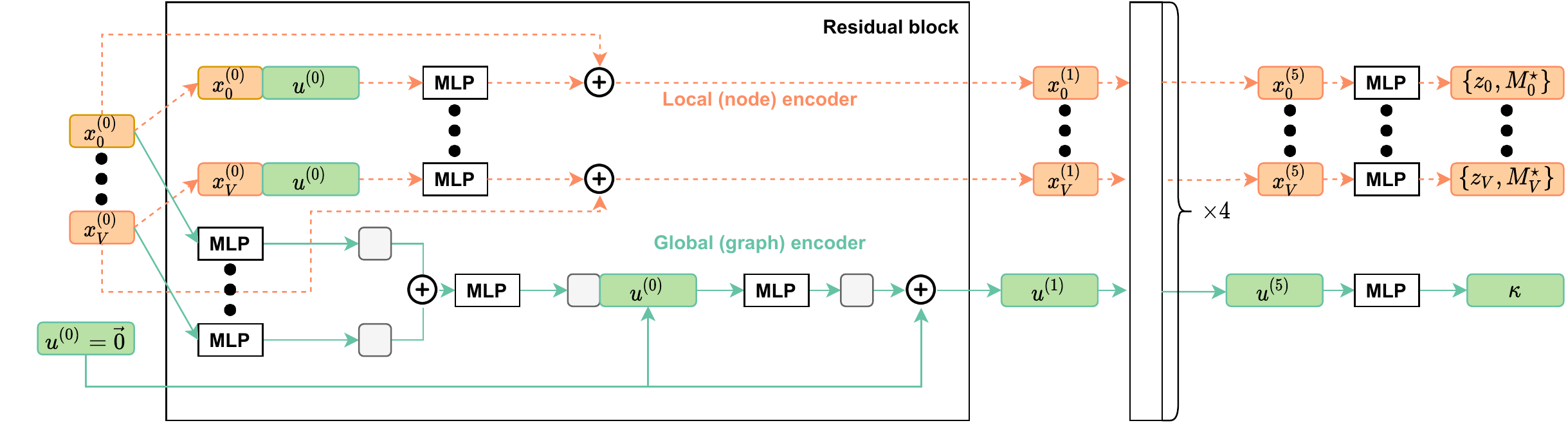}
\caption{Our network takes as input the features of $V$ nodes comprising a line of sight, denoted $\{x_0^{(0)}, \cdots, x_V^{(0)}\}$, and passes through five residual blocks. In each residual block $b$, the local (node) encodings $\{x_0^{(b-1)}, \cdots, x_V^{(b-1)}\}$ and global (graph) encoding $u^{(b-1)}$ are processed together through a series of residual MLPs to yield updated local and global encodings. The final encodings enter the projector MLP, which produces the posterior PDFs over the local target quantities ($z, M$ for each node) and the global target quantity $\kappa$.}
\label{fig:network_architecture}
\end{figure*}

A diagram of our network architecture is shown in Figure~\ref{fig:network_architecture}. Our network takes as input the features of $V$ nodes comprising a line of sight, denoted $\{x_0^{(0)}, \cdots, x_V^{(0)}\}$, and passes them through five residual blocks. In each residual block, indexed $b$, the local (node-level) encodings $\{x_0^{(b-1)}, \cdots, x_V^{(b-1)}\}$ and global (graph-level) encoding $u^{(b-1)}$ are processed together through a series of multi-layer perceptrons (MLPs) with residual skip connections to yield updated local and global encodings. The final encodings enter the last MLP, which produces the posterior PDFs over the local target quantities ($z, M_\star$ for each node) and the global target quantity $\kappa$. 

\subsection{Summary statistics matching} \label{sec:summary_stats}

Matching summary statistics can be considered an ABC-based method, whereby summary statistics of a test sightline are matched against those of training sightlines with known $\kappa$, which define the prior \citep{birrer2019h0licow}. We can compare the BGNN constraints with constraints produced by matching two types of summary statistics. One is the unweighted number counts ($N_{\rm sim}$), defined simply as the number of galaxies observed around the sightline. The other is the inverse-distance-weighted number counts ($N_{1/r, {\rm sim}}$). For $N_{\rm sim}$ observed galaxies in a field of view, 
\begin{align} \label{eq:n_inv_dist_def}
N_{1/r, {\rm sim}} \equiv \sum_{i=1}^N {\frac{1}{r_i + 10^{-5}}},
\end{align}
where $r_i$ is the projected flat-sky distance of galaxy $i$ from the central line of sight in arcseconds. The small value of $10^{-5}$ was added to the denominator for numerical stability. \chris{The subscript ``sim'' in $N_{\rm sim}$ and $N_{1/r, {\rm sim}}$ indicate that these are simplified statistics compared to the ones employed by TDCOSMO \citep[e.g.][]{greene2013improving, rusu2017h0licow}.} We did not consider more complex, model-dependent weighting schemes that incorporate inferred stellar masses and spectroscopic redshifts, e.g. $z, L, M$ in \cite{greene2013improving}, in order to focus on those that could be computed directly from photometric observations. We also did not normalize the summary statistics with respect to the average in the simulations, because all the sightlines in this study were derived from the same simulations. 

The summary statistics matching was implemented as follows. The values of $N_{\rm sim}$ ($N_{1/r, {\rm sim}}$) were computed for all the sightlines in the training and test sets. Then, for each test sightline, we queried the training sightlines with $N_{\rm sim}$ ($N_{1/r, {\rm sim}}$) that matched \liz{the test sightline's own} $N_{\rm sim}$ ($N_{1/r, {\rm sim}}$) within some closeness threshold, which we denote $\Delta N_{\rm sim}$ ($\Delta N_{1/r, {\rm sim}}$). The $\kappa$ corresponding to the matched training sightlines could then be interpreted as a posterior on $\kappa$, i.e. $p(\kappa|{\bm d}, \Omega_{\rm train})$. The threshold $\Delta N_{\rm sim}$ ($\Delta N_{1/r, {\rm sim}}$) determines how much information is captured in the matched samples. \liz{As a rule of thumb, it should be kept as small as possible provided that it yields sufficient matched samples. A large threshold yields more samples, but the quality of the matching may be poor. A small threshold sacrifices the number of samples for a closer match to the test sightline's summary statistic and thus potentially its $\kappa$. At one extreme, $\Delta N_{\rm sim}=0$ means that a training sample must have the same exact $N_{\rm sim}$ as the test sample in order to be considered ``matched'' and contribute to the posterior.} For a given test sightline, $\Delta N_{\rm sim}$ was chosen to be the \liz{smallest among $\{0, 1, 2, 4\}$ that yielded more than 100 matched samples. Likewise, $N_{1/r, {\rm sim}}$ was chosen to be the smallest among $\{1, 2, 4, 8\}$ that yielded more than 100 matched samples.} This resulted in $\Delta N_{\rm sim}=1$ for most sightlines and an average $\Delta N_{1/r, {\rm sim}}$ of 1.3. 

\chris{We stress that our matching scheme differs from that implemented by \cite{greene2013improving} and \cite{rusu2017h0licow} in two main ways. First, these studies binned the range of each summary statistic and reweighted the $\kappa$ contribution of each matched sightline in the final posterior by the inverse of the bin count, so that each bin contributed equal weight. The reweighting was intended to prevent the $\kappa$ posterior from being dominated by a large number of sightlines in a given $N_{\rm sim}$ bin. We do not enforce such a reweighting with respect to the summary statistic, but reweight by the training distribution in $\kappa$ when we hierarchically infer the hyperparameters (see Equation \ref{eq:omega_post}). Second, the $N_{\rm sim}$ and $N_{1/r, {\rm sim}}$ were matched jointly, rather than independently. We chose not to apply the joint constraint, because our small training set size of 200,000 sightlines made it difficult to achieve sufficient sample statistics. \cite{rusu2017h0licow} used a joint constraint from four types of summary statistics on $10^9$ training sightlines and note that using more than four may result in sample sparsity.} 

The $\kappa$ values of the matched training sightlines can be interpreted as posterior samples. We denote the matched samples and the posterior PDFs represented by these samples as 
\begin{align} \label{eq:posterior_N}
    \kappa_{{\rm matched}, N} \sim p_{N}\left( \kappa|\bm{d}, \Omega_{\rm train} \right)
\end{align}
and
\begin{align} \label{eq:posterior_N_inv_dist}
    \kappa_{{\rm matched}, {N_{1/r}}} \sim p_{N_{1/r}} \left( \kappa|\bm{d}, \Omega_{\rm train} \right),
\end{align}
for consistency with the notation of the BGNN posterior in Equation \ref{eq:predictive_distribution_variational}. We condition on $\Omega_{\rm train}$ everywhere, because both the BGNN and summary statistics techniques depend on the distributions of sightlines in the training set. 

\subsection{Hierarchical Inference} \label{sec:hierarchical_inference}

We designed eight test sets, each containing 1,000 sightlines drawn from normal distributions with varying $\mu$ and $\sigma$ (see Table \ref{tab:experiments}). They included shifted test sets with particularly overdense (high-$\kappa$) sightlines, where strong lenses tend to occur (B3, C3, C4). \chris{In particular, C4 serves as an extreme stress test, as it has very little overlap with the training $\kappa$ distribution.} See Figure \ref{fig:test_set_viz} for a visualization of the eight test distributions and their varying degrees of overlap with the training distribution. \chris{The size of 1,000 was chosen, because it was large enough to avoid small sample statistics but allowed experimental turnaround within reasonable time (see Section \ref{sec:computation_time}).}. 

The $\kappa$ posteriors in Equations \ref{eq:predictive_distribution_variational}, \ref{eq:posterior_N}, and \ref{eq:posterior_N_inv_dist} are conditioned on $\Omega_{\rm train}$, the set of assumptions governing the training set. Here, $\Omega_{\rm train}$ includes the particular choices of the underlying N-body simulation, semi-analytic models used to paint galaxies onto the halos, and the observation noise in the galaxy positions and magnitudes. The test set, in principle, may differ in any of these aspects. In this paper, we focus on training-test mismatches in the $\kappa$ distributions, by designing test sets from the same simulation suite as the training set but subsampled to follow shifted distributions in $\kappa$. As described in Section \ref{sec:simulated_data}, we assume Gaussian distributions for both the training and test distributions. 

We hierarchically infer the hyperparameters governing the Gaussian test distribution, $\mu_{\rm test}, \sigma_{\rm test}$, using importance sampling \citep{hogg2010inferring, foreman2014exoplanet, wagner2020hierarchical}. The posterior on $\mu_{\rm test}, \sigma_{\rm test}$ takes the form: 
\begin{align} \label{eq:omega_post}
    & p(\mu_{\rm test}, \log \sigma_{\rm test} | \{\bm{d}\}) \propto  \ p(\mu_{\rm test}, \log \sigma_{\rm test}) \times \nonumber \\
    & \prod_{i=1}^{N_{\rm test}} \frac{1}{M} \sum_{\kappa \sim p(\cdot | \bm{d}^{(i)}, \Omega_{\rm train})} \frac{p(\kappa|\mu_{\rm test}, \log \sigma_{\rm test})}{p(\kappa|\Omega_{\rm train})},
\end{align}
where $p(\mu_{\rm test}, \log \sigma_{\rm test})$ is our hyperprior and $i$ indexes the test set. Note that we work in the natural log space of $\sigma_{\rm test}$ to restrict our sampling to positive values. \adriano{See, e.g. \cite{wagner2020hierarchical}, for the derivation of Equation \ref{eq:omega_post}.} The second term lends itself to Markov Chain Monte Carlo (MCMC) sampling. To evaluate this MCMC objective for a candidate $\mu_{\rm test}, \sigma_{\rm test}$, we first draw $M$ samples from the individual posteriors conditioned on the training set. We can do this efficiently with a BGNN or use the matched samples themselves with the summary statistics method. We then evaluate the density of the training prior $p(\kappa|\Omega_{\rm train})$ and our target Gaussian distribution at those $\kappa$ samples and take the mean of the density ratio across samples. The weighting by $1/p(\kappa|\Omega_{\rm train})$ assigns more relative importance to $\kappa$ underrepresented by the training set. \chris{Note that division by zero will result whenever $p(\kappa |\mu_{\rm test}, \sigma_{\rm test})$ assigns density to a $\kappa$ value not encompassed by $p(\kappa|\Omega_{\rm train})$. We choose a Gaussian distribution, with support in $\mathbb{R}$, as the training prior, and ensure that it is sufficiently broad to make evaluation of the MCMC objective numerically stable.} Because Equation \ref{eq:omega_post} is not guaranteed to be unbiased for finite $M$, we carried out convergence tests on the hierarchical results to determine our $M$=20,000. This is a conservative value, given that there are only two hyperparameters. \adriano{We summarize the model (hyper)parameters and their (hyper)priors in Table \ref{tab:model_params}.}

We interpret the output of the BGNN and summary statistics matching as posterior PDFs, with the training prior applied. For evaluating $\kappa$ recovery on individual sightlines, we work with likelihoods by dividing by the training prior:
\begin{align} \label{eq:likelihoods}
    p(\bm{d}|\kappa, \Omega_{\rm train}) \propto p(\kappa | \bm{d},  \Omega_{\rm train})/p(\kappa | \Omega_{\rm train})
\end{align}
where $p(\kappa|\Omega_{\rm train}) = \mathcal{N}(0.01, 0.04)$ in our setup.

\begin{table*} 
\centering
\begin{tabular}{l|cc}
    \toprule
    {Parameter} & {Prior} & {Description}  \\
      \midrule
     \textbf{Individual lines of sight} & &  \\
     {$ \kappa$} & {$\mathcal{N}(0.01, 0.04)$} & {Posterior PDFs estimated by BGNN or summary statistics matching} \\
     \midrule
     \textbf{Gaussian test population} & & \\
  {$\mu_{\rm test}$} & {$\mathcal{U}(-0.5, 0.5)$} & {Gaussian mean} \\
  {$\log \sigma_{\rm test}$} & {$\mathcal{U}(-7, 0)$} & {Natural log of the Gaussian standard deviation} \\
    \bottomrule
  \end{tabular}
  \caption{Summary of model parameters}
  \label{tab:model_params}
\end{table*}%

\section{Results and Discussion}
\label{sec:results}
We organize our results as follows. In Section \ref{sec:individual_kappa}, we evaluate the BGNN's $\kappa$ recovery performance on individual sightlines. Having established that the individual constraints are reasonable, we proceed to present the hierarchical inference results of the test population as a whole in Section \ref{sec:population_kappa}. 

As summarized in Table \ref{tab:experiments}, our experiments vary in the mean ($\mu_{\rm test}$) and standard deviation ($\sigma_{\rm test}$) of the test $\kappa$ populations. Test set A is representative of the training set, i.e. follows the same distribution in $\kappa$. The four test sets in \chris{block C} share an artificially narrow $\sigma_{\rm test}$ of 0.005 but vary in the centers $\mu_{\rm test}$. The three test sets in \chris{block B} share a broader $\sigma_{\rm test}$ of 0.02 and vary in the $\mu_{\rm test}$. The resulting $\kappa$ distributions are visualized in Figure \ref{fig:test_set_viz}.

\begin{table} 
\begin{center}
\begin{tabular}{l|ccc}
    \toprule
    {Label} & {$\mu_{\rm test}$} &  {$\sigma_{\rm test}$} & {Description}\\
      \midrule
   A & {0.01} & {0.04} & {Representative of training set}\\
   \midrule
   B1 &  {-0.02} & {0.02} & {Underdense, broad} \\
  B2 & {0.0} & {0.02} & {Average, broad} \\
  B3 & {0.04} & {0.02} & {Overdense, broad} \\
     \midrule
     C1 & {-0.04} & {0.005} & {Underdense, narrow} \\
   C2 & {0} & {0.005} & {Average, narrow} \\
   C3 & {0.04} & {0.005} & {Overdense, narrow} \\
   C4 & {0.08} & {0.005} & {Extremely overdense, narrow} \\
    \bottomrule
  \end{tabular}
\end{center}
\caption{Summary of test sets}
  Summary of test sets defined by varying the input $\mu_{\rm test}, \sigma_{\rm test}$ All of the test sets had $N$=1,000 sightlines. \label{tab:experiments}
\end{table}

\subsection{Individual {$\kappa$} recovery}
\label{sec:individual_kappa}

\begin{figure*}[!htb]
\includegraphics[width=1\textwidth]{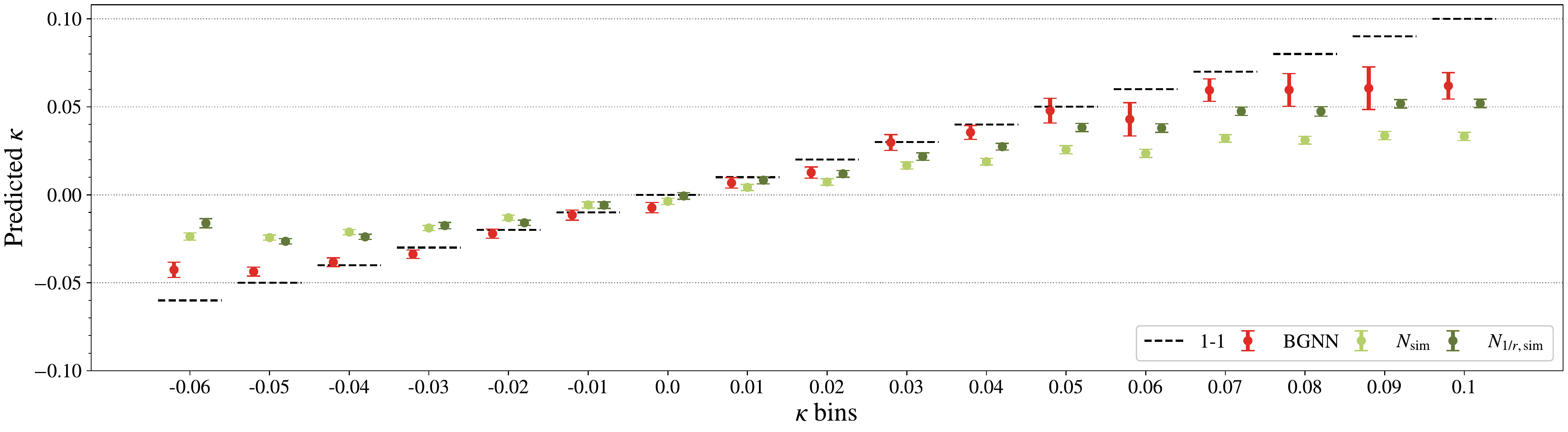}
\caption{The binned $\kappa$ recovery based on the likelihoods modeled by BGNN and the two summary statistics. The dots are computed from Equation \ref{eq:weighted_mean_bin} and the sizes of the error bars from Equation \ref{eq:weighted_std_bin}. The BGNN is more accurate than both $N_{\rm sim}$ and $N_{1/r, {\rm sim}}$ in all $\kappa$ bins. It boasts the highest accuracy in regions of $\kappa$ best represented in the training distribution $\sim \mathcal{N}(0.01, 0.04)$. On the other hand, all the methods reveal signs of upward bias in $\kappa < -0.05$ and downward bias in $\kappa > 0.06$. The bias is the worst for $N_{\rm sim}$, followed by $N_{1/r, {\rm sim}}$ and the BGNN. The bin count was 148 for all bins. \chris{Note that $N_{\rm sim}$ and $N_{1/r, {\rm sim}}$ are simplified versions of the summary statistics employed by TDCOSMO \citep[e.g.][]{rusu2017h0licow}, and our analysis operates on much fewer training sightlines ($2\times10^5$ compared to TDCOSMO's $10^9$).}}
\label{fig:binned_recovery_perbin}
\end{figure*}

Let us first compare the accuracy of $\kappa$ recovery across the BGNN and summary statistics methods for a range of $\kappa$ spanning the training distribution. In Figure \ref{fig:binned_recovery_perbin}, we bin the representative test set A into 17 $\kappa$ bins of fixed bin width 0.01 and plot \adriano{the mean and standard deviation} of the recovered $\kappa$ in each bin, weighted by the estimated uncertainties for the individual sightlines. That is, given the central estimate ($\kappa_i$) and the associated $1\sigma$ uncertainty ($\sigma_i$) of the likelihoods for each sightline $i$, we computed the weighted mean and standard error (SE) using $1/\sigma_i^2$ as the weights. Explicitly, we computed for each bin:
\begin{align} \label{eq:weighted_mean_bin}
    \textrm{weighted mean} \equiv {\bar \kappa} = \sum_{i=1}^B \frac{\kappa_i}{\sigma_i^2} \bigg/ W
\end{align}
and
\begin{align} \label{eq:weighted_std_bin}
    \textrm{weighted SE} = \sqrt{\sum_{i=1}^B \frac{(\kappa_i - \bar \kappa)^2}{\sigma_i^2} \times \frac{V}{W}},
\end{align}
where $B$ was the total number of sightlines in the bin, \chris{$W \equiv \sum_{i=1}^B 1/\sigma_i^2$} was the sum of the weights, and the $V \equiv W_2/W^2$ where $W_2 \equiv \sum_{i=1}^B 1/\sigma_i^4$. If all weights $1/\sigma_i^2$ are equal, the factor $V$ reduces to $1/B$, so Equation \ref{eq:weighted_std_bin} becomes the conventional (unweighted) SE that scales with $1/\sqrt{B}$. In Figure \ref{fig:binned_recovery_perbin}, the dots are computed from Equation \ref{eq:weighted_mean_bin} and the sizes of the error bars from Equation \ref{eq:weighted_std_bin}. Overall, our BGNN is more accurate than both $N_{\rm sim}$ and $N_{1/r, {\rm sim}}$ in all $\kappa$ bins. It boasts the highest accuracy in regions of $\kappa$ best represented in the training distribution $\sim \mathcal{N}(0.01, 0.04)$. On the other hand, all the methods reveal signs of upward bias in $\kappa < -0.05$ and downward bias in $\kappa > 0.06$ given the standard errors per bin. The bias is the worst for $N_{\rm sim}$, followed by $N_{1/r, {\rm sim}}$ and the BGNN.

\begin{figure*}[!htb]
\includegraphics[width=1\textwidth]{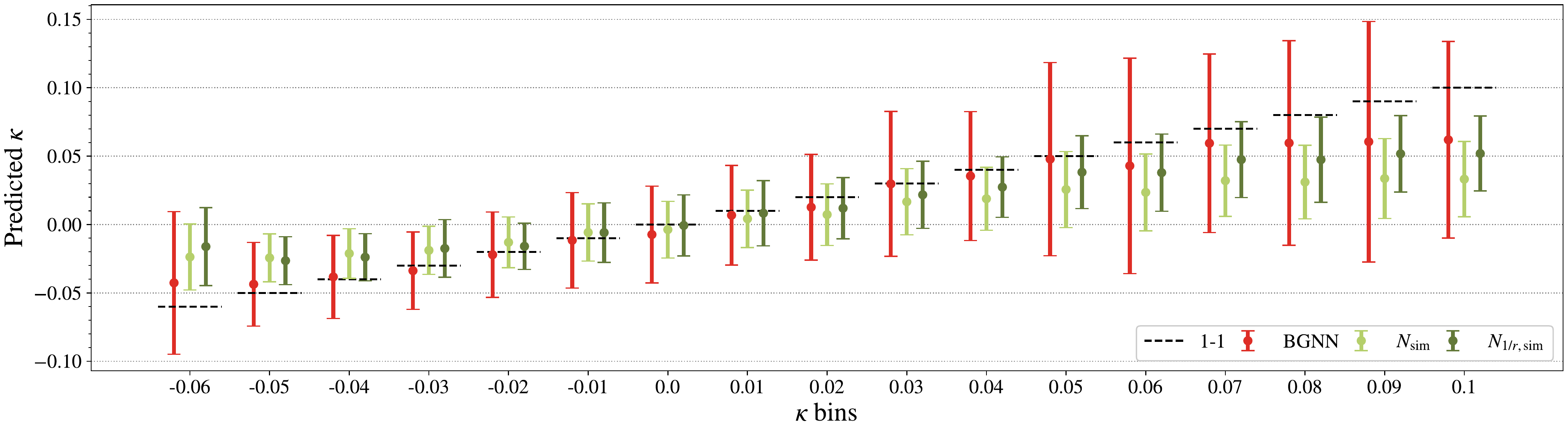}
\caption{The binned $\kappa$ recovery based on the likelihoods modeled by BGNN and the two summary statistics. The dots are computed from Equation \ref{eq:weighted_mean_bin} and the sizes of the error bars from Equation \ref{eq:weighted_std_bin_per_kappa}. The BGNN predictions are consistent with the true $\kappa$ for all $\kappa$ values. On the other hand, both the $N_{\rm sim}$ and $N_{1/r, {\rm sim}}$ summary statistics reveal signs of upward bias in $\kappa < -0.04$ and downward bias in $\kappa > 0.09$. The bin count was 148 for all bins. \chris{Note that $N_{\rm sim}$ and $N_{1/r, {\rm sim}}$ are simplified versions of the summary statistics employed by TDCOSMO \citep[e.g.][]{rusu2017h0licow}, and our analysis operates on much fewer training sightlines ($2\times10^5$ compared to TDCOSMO's $10^9$).}}
\label{fig:binned_recovery_perkappa}
\end{figure*}

Also interesting is the level of accuracy for each value of $\kappa$, rather than for each bin. To assess this, we apply a correction factor $K \equiv 1/(1 - V)$ \citep{bevington2003data} to account for the varying bin count, i.e.
\begin{align} \label{eq:weighted_std_bin_per_kappa}
    \textrm{weighted spread per $\kappa$} = \sqrt{ \sum_{i=1}^B \frac{(\kappa_i - \bar \kappa)^2}{\sigma_i^2} \times \frac{K}{W}}.
\end{align}
\liz{The error bars in Figure \ref{fig:binned_recovery_perkappa} are computed from Equation \ref{eq:weighted_std_bin_per_kappa} whereas the dots are the same as in Figure \ref{fig:binned_recovery_perbin}.} We find that, on a per-$\kappa$ (per-sightline) level, both the $N_{\rm sim}$ and $N_{1/r, {\rm sim}}$ summary statistics reveal signs of upward bias in $\kappa < -0.04$ and downward bias in $\kappa > 0.09$. The BGNN predictions are consistent with the true $\kappa$ for all $\kappa$ values. 

We choose three metrics to quantitatively compare the $\kappa$ inference performance of the BGNN and the summary statistics matching across experiments. The first metric is the log likelihood evaluated at the truth:
\begin{equation} \label{eq:logp}
{\log p} \equiv \log p(\bm{d}| \kappa, \Omega_{\rm train})|_{\kappa = \kappa_{\rm true}},
\end{equation}
i.e. the inferred likelihood from Equation \ref{eq:likelihoods} evaluated at a sightline's true $\kappa$ value. Higher $\log p$ is more accurate and/or more precise. We also use the Median Absolute Error (MAE) and Median Absolute Deviation (MAD) as robust measures of accuracy and precision, respectively, with smaller values being better. These are given by:
\begin{align} \label{eq:mae}
{\rm MAE} &\equiv \textsc{median}( | \kappa_{\rm sample} - \kappa_{\rm true}| ) \\
\label{eq:mad}
{\rm MAD} &\equiv \textsc{median}(| \kappa_{\rm sample} - \textsc{median}(\kappa_{\rm sample})| )/1.4826
\end{align}
where $\kappa_{\rm sample} \sim p(\bm{d}|\kappa, \Omega_{\rm train})$ for a given sightline with true convergence $\kappa_{\rm true}$. \chris{The factor of 1/1.4826 converts the MAD of a Gaussian into the standard deviation}, so we can interpret Equation \ref{eq:mad} as a robust measure of standard deviation. Note that for the BGNN experiments, $k_{\rm samples}$ constitute samples from the BGNN likelihood for this sightline, while for the summary statistics methods, $k_{\rm samples}$ are the matched training set samples reweighted according to Equation \ref{eq:likelihoods}. 

Table \ref{tab:individual_metrics} compares these metrics across BGNN, $N_{\rm sim}$, and $N_{1/r, {\rm sim}}$ for a grid of $\kappa$ bins. Values listed are the median and standard deviation taken across all the sightlines in each bin. The BGNN is generally more accurate and more precise compared to either summary statistic, in terms of $\log p$, MAE, and MAD. Note, however, that these metrics weight each sightline equally. In order to account for the varying information content across the sightlines, we examine the hierarchical inference results of population $\kappa$ statistics in the next section.

\begin{table*}[ht!]
\setlength{\extrarowheight}{3pt}
\small
  \caption{Metrics Evaluating $\kappa$ Recovery on Individual Test Sightlines for Each $\kappa$ Bin}
  \centering
  \begin{tabular}{K|KLLL||K|KLLL}
    \toprule
    $\kappa$     & Method   & $\log p^{ab}$ & MAE$^{ac}$ (Accuracy) & MAD$^{ad}$ (Precision) & $\kappa$ & Method   & $\log p^{ab}$ & MAE$^{ac}$ (Accuracy) & MAD$^{ad}$ (Precision)\\
    \midrule
{[-0.06, -0.04)} & {BGNN}  & 2.6 $\pm$ 0.3  & 0.02 $\pm$ 0.01  &
0.03 $\pm$ 0.01
 &
{[0.02, 0.04)} & {BGNN}  & 1.9 $\pm$ 0.2  & 0.04 $\pm$ 0.01  &
0.03 $\pm$ 0.01
\\
{}  & {$N_{\rm sim}$}  & 2.3 $\pm$ 0.4  & 0.03 $\pm$ 0.01  &
0.03 $\pm$ 0.01
 &
{}  & {$N_{\rm sim}$}  & 1.3 $\pm$ 0.4  & 0.07 $\pm$ 0.02  &
0.05 $\pm$ 0.01
\\
{}  & {$N_{1/r, {\rm sim}}$}  & 2.1 $\pm$ 0.2  & 0.03 $\pm$ 0.01  &
0.02 $\pm$ 0.01
 &
{}  & {$N_{1/r, {\rm sim}}$}  & 1.6 $\pm$ 0.4  & 0.05 $\pm$ 0.02  &
0.04 $\pm$ 0.01
\\
\hline
{[-0.04, -0.02)} & {BGNN}  & 2.4 $\pm$ 0.2  & 0.02 $\pm$ 0.01  &
0.03 $\pm$ 0.01
 &
{[0.04, 0.06)} & {BGNN}  & 1.8 $\pm$ 0.2  & 0.04 $\pm$ 0.01  &
0.03 $\pm$ 0.01
\\
{}  & {$N_{\rm sim}$}  & 2.1 $\pm$ 0.3  & 0.03 $\pm$ 0.01  &
0.03 $\pm$ 0.01
 &
{}  & {$N_{\rm sim}$}  & 1.3 $\pm$ 0.3  & 0.07 $\pm$ 0.02  &
0.05 $\pm$ 0.01
\\
{}  & {$N_{1/r, {\rm sim}}$}  & 2.0 $\pm$ 0.1  & 0.03 $\pm$ 0.01  &
0.02 $\pm$ 0.01
 &
{}  & {$N_{1/r, {\rm sim}}$}  & 1.4 $\pm$ 0.4  & 0.06 $\pm$ 0.02  &
0.04 $\pm$ 0.01
\\
\hline
{[-0.02, 0.0)} & {BGNN}  & 2.2 $\pm$ 0.2  & 0.03 $\pm$ 0.01  &
0.03 $\pm$ 0.01
 &
{[0.06, 0.08)} & {BGNN}  & 1.7 $\pm$ 0.4  & 0.05 $\pm$ 0.01  &
0.03 $\pm$ 0.01
\\
{}  & {$N_{\rm sim}$}  & 1.8 $\pm$ 0.3  & 0.05 $\pm$ 0.01  &
0.04 $\pm$ 0.01
 &
{}  & {$N_{\rm sim}$}  & 1.3 $\pm$ 0.4  & 0.06 $\pm$ 0.02  &
0.05 $\pm$ 0.01
\\
{}  & {$N_{1/r, {\rm sim}}$}  & 2.0 $\pm$ 0.1  & 0.03 $\pm$ 0.01  &
0.03 $\pm$ 0.01
 &
{}  & {$N_{1/r, {\rm sim}}$}  & 1.3 $\pm$ 0.4  & 0.06 $\pm$ 0.02  &
0.04 $\pm$ 0.01
\\
\hline
{[0.0, 0.02)} & {BGNN}  & 2.0 $\pm$ 0.2  & 0.03 $\pm$ 0.01  &
0.03 $\pm$ 0.01
 &
{[0.08, 0.1)} & {BGNN}  & 1.6 $\pm$ 0.5  & 0.05 $\pm$ 0.02  &
0.03 $\pm$ 0.01
\\
{}  & {$N_{\rm sim}$}  & 1.6 $\pm$ 0.4  & 0.05 $\pm$ 0.02  &
0.04 $\pm$ 0.01
 &
{}  & {$N_{\rm sim}$}  & 1.2 $\pm$ 0.5  & 0.07 $\pm$ 0.03  &
0.05 $\pm$ 0.01
\\
{}  & {$N_{1/r, {\rm sim}}$}  & 1.9 $\pm$ 0.3  & 0.04 $\pm$ 0.01  &
0.03 $\pm$ 0.01
 &
{}  & {$N_{1/r, {\rm sim}}$}  & 1.4 $\pm$ 0.4  & 0.06 $\pm$ 0.02  &
0.04 $\pm$ 0.01
\\
\hline
{all} & {BGNN} & {1.9} & {0.03} & {0.03} & {} & {} & {} & {}& {} \\
{} & {$N_{\rm sim}$} & {1.6} & {0.05} & {0.04} & {} & {} & {} & {}& {} \\
{} & {$N_{1/r, {\rm sim}}$} & {1.8} & {0.04} & {0.03} & {} & {} & {} & {}& {} \\
    \bottomrule
  \end{tabular}
  \label{tab:individual_metrics}
\begin{flushleft}
{\footnotesize{$^a$ Values listed are the median of the metrics across sightlines in a given bin. Errors listed are the standard deviations in a given bin.}}\\
{\footnotesize{$^b$ Defined in Equation \ref{eq:logp}. Higher is better.}}\\
{\footnotesize{$^c$ Defined in Equation \ref{eq:mae}. {Lower is more accurate.}}}\\
{\footnotesize{$^d$ Defined in Equation \ref{eq:mad}. Lower is more precise.}}
\end{flushleft}
\vspace*{0.2cm}
\end{table*}
\normalsize

\subsection{Population inference}
\label{sec:population_kappa}

We proceed to present the hierarchical inference results on the population $\kappa$ statistics. \liz{Recall that our experiments vary in the mean $\mu_{\rm test}$ and standard deviation $\sigma_{\rm test}$ of the test populations. Here we compare the BGNN's recovery performance on our test sets to understand the sensitivity of the model across the hyperparameter space.} The simple and analytic $H_0$ error decomposition by \cite{birrer2021hubble} described in Section \ref{sec:cosmography} also allows us to connect the precision and accuracy of $\mu_{\rm test}$ recovery to the $H_0$ error budget. 

Table \ref{tab:hier_metrics_cosmo} summarizes the bias and uncertainty of $\mu_{\rm test}$ constraints for all the experiments. Treating the sightlines in each experiment as a population of strong lensing sightlines, we can interpret our maximum-likelihood estimate of $\mu_{\rm test}$ as the central estimate of the external convergence, $\bar \kappa_{\rm test}$, and the uncertainty on $\mu_{\rm test}$ as $\sigma(\bar \kappa_{\rm ext})$. Equation \ref{eq:mean_contrib_H0} then allows us to compute the contribution of the external environment to the overall fractional uncertainty on $H_0$. 

For all experiments, the BGNN introduces lower bias on $H_0$ (last column) than does either summary statistic. With the exception of the extremely overdense population C4, which introduces a 1.6\% bias on $H_0$, the bias level remains at sub-percent for the BGNN.

The performances of all three methods follow patterns expected from the individual recovery results in Section \ref{sec:individual_kappa}---the best constraints on $\mu_{\rm test}$ come from environments best represented by the training set, i.e. B2, C2 that have $\kappa \sim 0$. Notably, the percent bias on $H_0$ is only -0.1\% on B2 and -0.2\% on C2 for the BGNN. 

The bias increases towards underdense and overdense populations in the tails of the training distribution, but the BGNN is significantly more robust than the summary statistics. Whereas the bias level only increased to ${\sim}0.8\%$ from C2 to C1/C3 for the BGNN, it increased to ${\sim}2\%$  for $N_{\rm sim}$ and ${\sim}1.5\%$ for $N_{1/r, {\rm sim}}$. A similar pattern holds for the broad B group of test sets as well, although the degree of deterioration is slightly worse for the B group because it contains more of the extreme sightlines.

We find that the 1$\sigma$ uncertainty on $H_0$ (second-to-last column) is smaller than the bias on $H_0$ across the board, pointing to a global underestimation of the uncertainty on $\mu_{\rm test}$. The only exception is the BGNN for B2, where the bias is small enough (-0.1\%) that the estimated  1$\sigma$ uncertainty of 0.14\% can account for it. Even so, relative to the summary statistics, the BGNN estimates of the uncertainty are \adriano{better, i.e. comes closer to covering the bias.} 

\begin{table*}[ht!]
\setlength{\extrarowheight}{3pt}
\small
  \caption{Recovery of population $\kappa$ and impact on $H_0$ error budget}
  \centering
  \begin{tabular}{K|KMMMMM}
    \toprule
    {} & {} & {Central $\mu_{\rm test}$ pred} & {Bias on $\mu_{\rm test}$} & {1$\sigma$ uncert. on $\mu_{\rm test}$} & {\% contrib to $H_0$ err$^a$} & {\% bias on $H_0^b$}
    \\
   Exp.     & Method  & {$\bar \kappa_{\rm ext} $} & $\bar \kappa_{\rm ext} - \mu_{\rm test, true}$ & $\sigma(\bar \kappa_{\rm ext})$ & {$\frac{\sigma(\bar \kappa_{\rm ext})}{1 - \bar \kappa_{\rm ext}}$} & {$\frac{ \mu_{\rm test, true} - \bar \kappa_{\rm ext}}{1 - \mu_{\rm test, true}}$} \\
    \midrule
\bf{ A } & BGNN & 0.006 & -0.004 & 0.002 & 0.157 & 0.417 \\
{} & $N_{\rm sim}$ & 0.002 & -0.008 & 0.002 & 0.183 & 0.826 \\
{} & $N_{1/r, {\rm sim}}$ & 0.005 & -0.005 & 0.002 & 0.188 & 0.485 \\
\hline
\hline
\bf{ B1 } & BGNN & -0.032 & 0.008 & 0.001 & 0.088 & -0.814 \\
{} & $N_{\rm sim}$ & -0.019 & 0.021 & 0.001 & 0.063 & -2.010 \\
{} & $N_{1/r, {\rm sim}}$ & -0.022 & 0.018 & 0.001 & 0.068 & -1.739 \\
\hline
\bf{ B2 } & BGNN & 0.001 & 0.001 & 0.001 & 0.143 & -0.108 \\
{} & $N_{\rm sim}$ & -0.004 & -0.004 & 0.001 & 0.116 & 0.399 \\
{} & $N_{1/r, {\rm sim}}$ & -0.003 & -0.003 & 0.001 & 0.104 & 0.323 \\
\hline
\bf{ B3 } & BGNN & 0.031 & -0.009 & 0.002 & 0.163 & 0.950 \\
{} & $N_{\rm sim}$ & 0.018 & -0.022 & 0.001 & 0.094 & 2.273 \\
{} & $N_{1/r, {\rm sim}}$ & 0.025 & -0.015 & 0.001 & 0.090 & 1.597 \\
\hline
\hline
\bf{ C1 } & BGNN & -0.032 & 0.008 & 0.001 & 0.120 & -0.762 \\
{} & $N_{\rm sim}$ & -0.021 & 0.019 & 0.001 & 0.053 & -1.866 \\
{} & $N_{1/r, {\rm sim}}$ & -0.024 & 0.016 & 0.001 & 0.060 & -1.528 \\
\hline
\bf{ C2 } & BGNN & 0.002 & 0.002 & 0.001 & 0.139 & -0.241 \\
{} & $N_{\rm sim}$ & -0.003 & -0.003 & 0.001 & 0.146 & 0.322 \\
{} & $N_{1/r, {\rm sim}}$ & -0.004 & -0.004 & 0.001 & 0.099 & 0.415 \\
\hline
\bf{ C3 } & BGNN & 0.033 & -0.007 & 0.002 & 0.164 & 0.750 \\
{} & $N_{\rm sim}$ & 0.020 & -0.020 & 0.001 & 0.094 & 2.125 \\
{} & $N_{1/r, {\rm sim}}$ & 0.027 & -0.013 & 0.001 & 0.084 & 1.404 \\
\hline
\bf{ C4 } & BGNN & 0.065 & -0.015 & 0.003 & 0.271 & 1.656 \\
{} & $N_{\rm sim}$ & 0.029 & -0.051 & 0.001 & 0.103 & 5.579 \\
{} & $N_{1/r, {\rm sim}}$ & 0.040 & -0.040 & 0.001 & 0.086 & 4.388 \\
\hline
    \bottomrule
  \end{tabular}
  \label{tab:hier_metrics_cosmo}
\begin{flushleft}
{\footnotesize{$^a$ Primary contribution to the fractional $H_0$ error budget based on $\mu_{\rm test}$ constraints, defined in Equation \ref{eq:mean_contrib_H0}.}}\\
{\footnotesize{$^b$ Bias introduced in $H_0$ due to bias on $\mu_{\rm test}$, defined in Equation \ref{eq:h0_bias}.}}\\
\end{flushleft}
\end{table*}
\normalsize

\begin{figure}[!htb]
\includegraphics[width=0.48\textwidth]{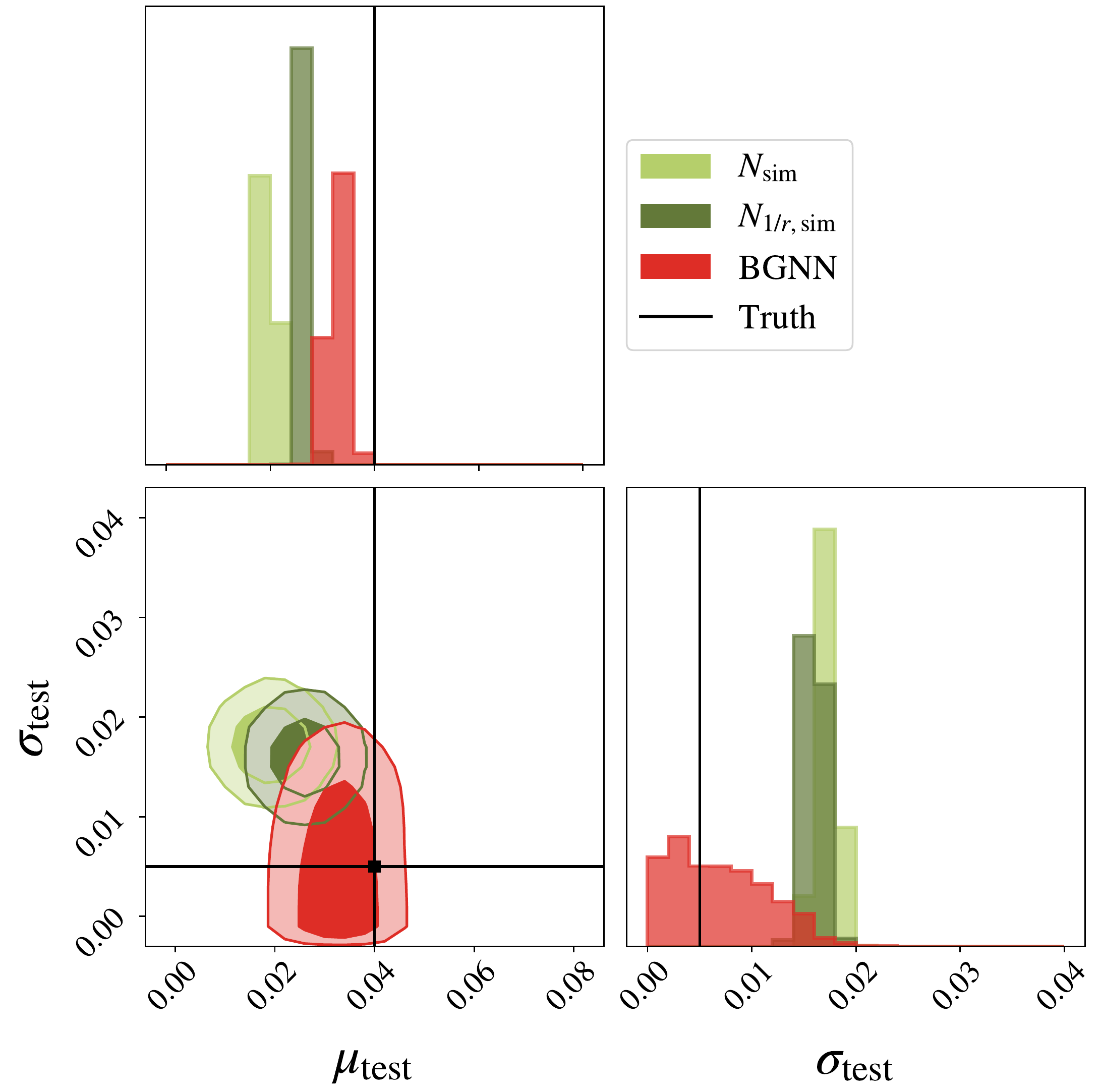}
\caption{Constraints on the population hyperparameters for the narrow, overdense test set (Experiment C3) with truth $\mu_{\rm test}=0.04$ and $\sigma_{\rm test}=0.005$. Contours are 68\% and 95\% credible intervals. BGNN can accurately recover the $\mu_{\rm test}, \sigma_{\rm test}$ within its 1$\sigma$ credible interval. Both the $N_{\rm sim}$ and $N_{1/r, {\rm sim}}$ summary statistics lead to downward biases on $\mu_{\rm test}$. They also overestimate $\sigma_{\rm test}$. \chris{Note that $N_{\rm sim}$ and $N_{1/r, {\rm sim}}$ are simplified versions of the summary statistics employed by TDCOSMO \citep[e.g.][]{rusu2017h0licow}, and our analysis operates on much fewer training sightlines ($2\times10^5$ compared to TDCOSMO's $10^9$).}}
\label{fig:hier_E4}
\end{figure}

The hierarchical inference constraints for C3 is shown in Figure \ref{fig:hier_E4}. This test set represents a particularly challenging environment with $\mu_{\rm test}$ shifted high and $\sigma_{\rm test}$ artificially narrow. Our BGNN can accurately recover the high $\mu_{\rm test}$ and narrow $\sigma_{\rm test}$ within its 1$\sigma$ credible interval. The BGNN constraints translate into a 0.3\% contribution to the $H_0$ uncertainty without evidence of bias. On the other hand, both the $N_{\rm sim}$ and $N_{1/r, {\rm sim}}$ summary statistics lead to downward biases on $\mu_{\rm test}$. They also overestimate $\sigma_{\rm test}$, which happens with underestimated uncertainties on individual $\kappa$. 

\begin{figure}[!htb]
\includegraphics[width=0.48\textwidth]{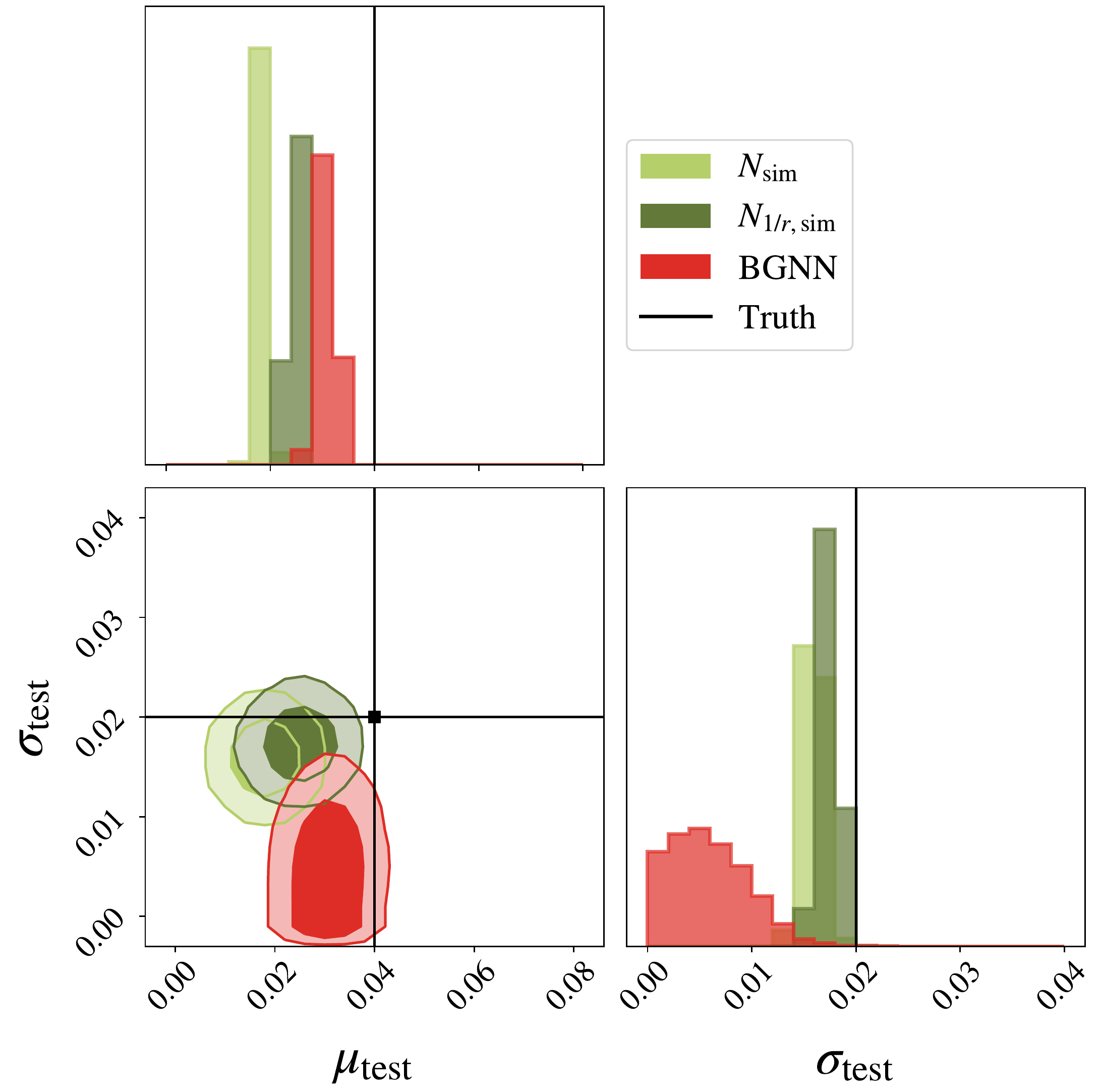}
\caption{Constraints on the population hyperparameters for the broad, overdense test set (Experiment B3) with truth $\mu_{\rm test}=0.04$ and $\sigma_{\rm test}=0.02$. Contours are 68\% and 95\% credible intervals. Accurate $\mu_{\rm test}$ recovery is critical to unbiased $H_0$ inference and the impact of $\sigma_{\rm test}$ recovery is second-order (Section \ref{sec:cosmography}). The BGNN can accurately recover the $\mu_{\rm test}, \sigma_{\rm test}$ within its 2$\sigma$ credible interval. Both the $N_{\rm sim}$ and $N_{1/r, {\rm sim}}$ summary statistics lead to downward biases on $\mu_{\rm test}$. \chris{Note that $N_{\rm sim}$ and $N_{1/r, {\rm sim}}$ are simplified versions of the summary statistics employed by TDCOSMO \citep[e.g.][]{rusu2017h0licow}, and our analysis operates on much fewer training sightlines ($2\times10^5$ compared to TDCOSMO's $10^9$).}}
\label{fig:hier_E8}
\end{figure}

Similarly, the constraints for the broad, overdense test set (Experiment B3) is shown in Figure \ref{fig:hier_E8}. Out of all our test distributions, B3 best characterizes the $\kappa_{\rm ext}$ distributions in the seven TDCOSMO lenses. Our BGNN can accurately recover the high $\mu_{\rm test}$ within its 2$\sigma$ credible interval. Again, both the summary statistics underestimate $\mu_{\rm test}$ and overestimate $\sigma_{\rm test}$. 

Our hierarchical inference pipeline reaches a numerical breaking point, however, with a test set shifted far away from the training set, e.g. Experiment C4 which lies at the upper tail of the training distribution. The BGNN struggles because it has not been exposed to many sightlines with $\kappa \sim 0.08$ in the training set. The low training density in this region of high $\kappa$ affects the summary statistics methods more directly, because they run out of training samples to match. See Figure \ref{fig:test_set_viz} for a visualization of the $\kappa$ distribution in Experiment C4 compared to that of the training set. 

Even at this breaking point, however, the true $\mu_{\rm test}$ falls within the BGNN's 2$\sigma$ credible interval. For the summary statistics, the downward biases on $\mu_{\rm test}$ become more dramatic than in Experiments C3 or B3. The uncertainties are underestimated for the BGNN in this regime, resulting in overestimation of $\sigma_{\rm test}$. \chris{The summary statistics methods do recover $\sigma_{\rm test}$ within their 2$\sigma$ credible interval, however.}

\begin{figure}[!htb]
\includegraphics[width=0.48\textwidth]{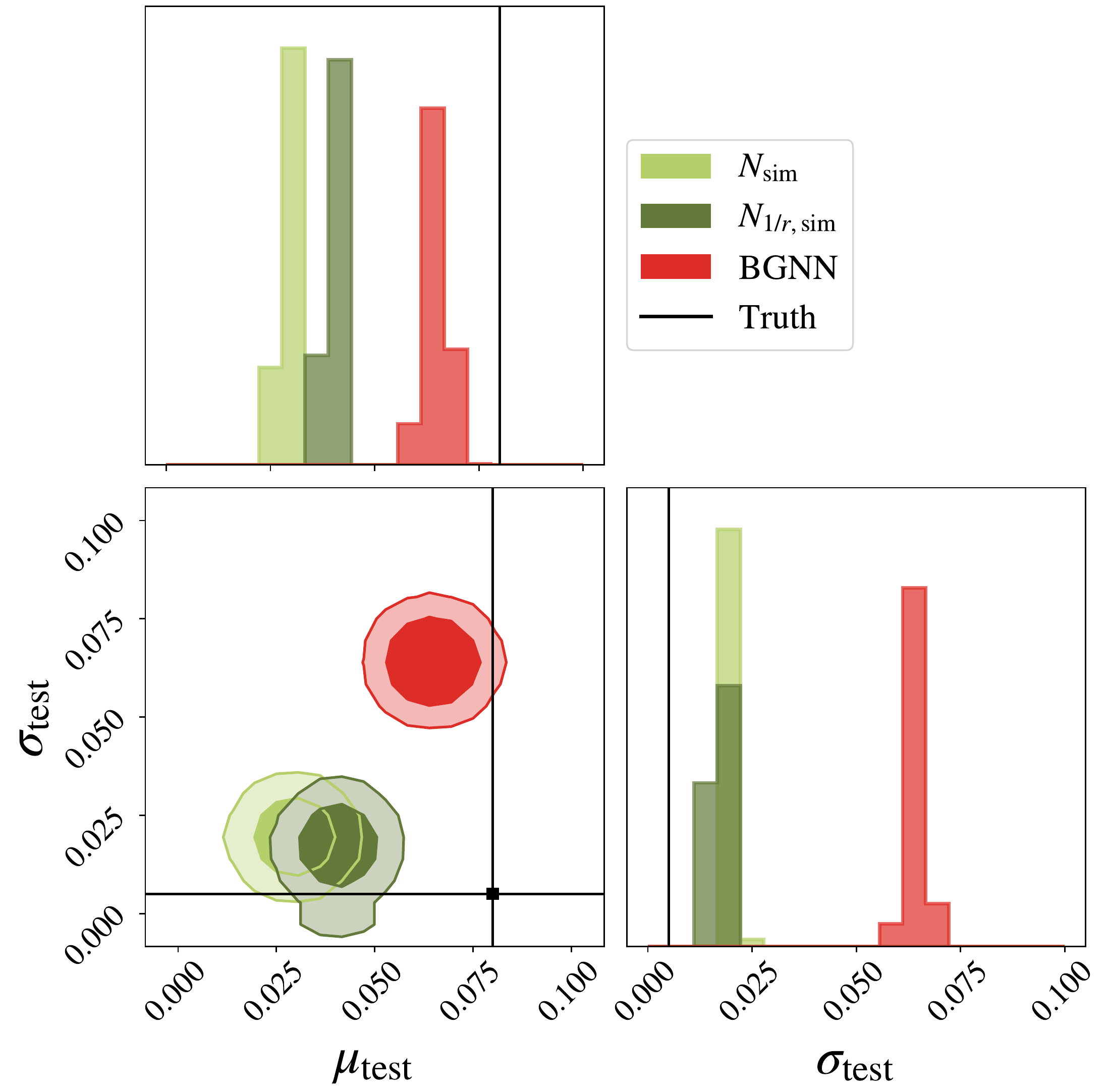}
\caption{Constraints on the population hyperparameters for the narrow, extremely overdense test set (Experiment C4) with truth $\mu_{\rm test}=0.08$ and $\sigma_{\rm test}=0.005$. Contours are 68\% and 95\% credible intervals. Accurate $\mu_{\rm test}$ recovery is critical to unbiased $H_0$ inference and the impact of $\sigma_{\rm test}$ recovery is second-order (Section \ref{sec:cosmography}). The BGNN can still recover the $\mu_{\rm test}$ within its 2$\sigma$ credible interval, but the $\sigma_{\rm test}$ is overestimated by a factor of 15. Both the $N_{\rm sim}$ and $N_{1/r, {\rm sim}}$ summary statistics lead to severe downward biases on $\mu_{\rm test}$ but do recover $\sigma_{\rm test}$ within their 2$\sigma$ credible interval.s. \chris{Note that $N_{\rm sim}$ and $N_{1/r, {\rm sim}}$ are simplified versions of the summary statistics employed by TDCOSMO \citep[e.g.][]{rusu2017h0licow}, and our analysis operates on much fewer training sightlines ($2\times10^5$ compared to TDCOSMO's $10^9$).}}
\label{fig:hier_E5}
\end{figure}

Based on a simulated dataset consisting of 200,000 sightlines following $\mathcal{N}(0.01, 0.04)$, we have demonstrated the ability of BGNNs to effectively model the mapping between photometric observations and the continuous $\kappa$ space. Summary statistics matching, on the other hand, operates on discrete samples so cannot extrapolate beyond the $\kappa$ samples present in the dataset. \chris{When there are only 200,000 sightlines, the poor matching causes posteriors to revert to the prior, and when the prior is shifted from the actual population, bias results. The BGNN also tends toward the training distribution when trained on a small dataset and also leads to bias, but it is more robust to the train-test mismatch.} For both approaches, we expect the hierarchical inference performance to improve with bigger training sets that better sample high- and low- $\kappa$ environments. Further tests are needed to probe the relative performance of the BGNN and summary statistics methods with increasing training sets. We postpone this exercise to future work.

\chris{We emphasize that the comparison between BGNN and summary statistics presented in this work is not to be viewed as an evaluation of the TDCOSMO analysis of $\kappa_{\rm ext}$. We include the comparison to highlight the difference between a neural net that operates on the continuous $\kappa$ space and an ABC method that either accepts or rejects discrete samples. Our implementation of the summary statistics matching differs in significant ways from the TDCOSMO implementation. TDCOSMO uses orders of magnitude more sightlines ($\sim$billion compared to our 200,000) \citep[e.g.][]{greene2013improving,rusu2017h0licow}. Their sightlines follow the phenomenological distribution from the Millennium Simulation which extends out to high $\kappa$. Thanks to the dataset size, they are able to match to both $N_{\rm sim}$ and $N_{1/r, {\rm sim}}$, whereas we have matched to either $N_{\rm sim}$ or $N_{1/r, {\rm sim}}$ in our work. Additionally, TDCOSMO incorporates external shear information from the lens modeling to further constrain their $\kappa_{\rm ext}$. This could not be done here, because we focused on sightlines without strong lensing.

A key takeaway from our toy hierarchical tests is the importance of systematically testing our assumptions about the $\kappa$ prior. Regardless of the $\kappa$ inference algorithm used, neural network or ABC-based, we must infer the target $\kappa$ population and account for any difference from the prior. If we instead derive $\kappa$ posteriors from individual sightlines based on a chosen $\kappa$ prior and let them independently enter the joint downstream $H_0$ analysis, we effectively multiply by the prior $N$ times, where $N$ is the number of sightlines. As $N$ increases, the particular choice of $\kappa$ prior will become more significant. 
}
\chris{
\subsection{Computation time}
\label{sec:computation_time}
The total computation time of our hierarchical $\kappa$ inference pipeline can be broken down into the BGNN training/inference time and the hiearchical inference time. Training the BGNN with the configuration detailed in Section \ref{sec:bgnn} took about 12 hours. The model was trained for 118 epochs on an Intel Xeon Gold 6148 GPU on the Cori system available at the National Energy Research Scientific Computing Center (NERSC). Generating BGNN predictions on the 1,000 sightlines in each test set took seconds. For a given training set size, the BGNN training and inference time can be considered fixed with respect to the number of test sightlines. The last step of hierarchical inference involves MCMC sampling, which took 2-3 hours to converge on 4 CPU cores for 1,000 sightlines. In our current implementation, the evaluation of the MCMC objective scales linearly with the number of sightlines in the test set.  

The summary statistics matching took 30 minutes for 1,000 sightlines on 4 CPU cores, for either $N_{\rm sim}$ or $N_{1/r, {\rm sim}}$. In our current implementation, the matching time scaled linearly with the number of sightlines and with the size of the ``closeness threshold grid.'' For each sightline, we carried out matching according to each threshold on this prespecified grid and took the smallest threshold with more than 100 matched samples at the end. We used a grid of size 20 for both $N_{\rm sim}$ and $N_{1/r, {\rm sim}}$. In the hierarchical inference step, the MCMC sampling with the summary statistics likelihoods also took 2-3 hours to converge on 4 CPU cores for 1,000 sightlines --- similarly as with the BGNN likelihoods.

To generate our datasets, we performed additional raytracing on top of the \textsc{CosmoDC2} values, as detailed in Section \ref{sec:structure_enhanced}, to compute the $\kappa$ labels and queried the galaxy catalog to construct the input. We simulated 850,000 sightlines and subsampled smaller Gaussian subsets to conduct our analysis. Out of the 850,000 sightlines, we reserved 500,000 sightlines for subsampling the Gaussian training set of 200,000 sightlines and used the remaining 250,000 sightlines to subsample the Gaussian validation and test sets. The raytracing took 85 hours for 850,000 sightlines on 18 cores. The input construction took 102 hours for the same sightlines. 
}
\section{Conclusion}
\label{sec:conclusion}

In this paper, we introduced a novel graph-based neural network architecture that can infer $\kappa$ from any number of observable galaxy properties and a first hierarchical pipeline for constraining the population $\kappa$ statistics. To probe a range of numerics in the recovery of hyperparameters in the target population, we have designed toy testbeds where the training $\kappa$ distribution (our prior) and test populations are mismatched to varying degrees. \chris{We present comparisons between BGNN and summary statistics matching throughout to showcase the distinction between a neural network that operates in a continuous $\kappa$ space and an ABC-based method that either accepts or reject discrete samples. Note, however, that our implementation of summary statistics matching does not represent TDCOSMO's, which is more advanced and factors in more information.}

We conclude that:
\begin{itemize}
    \item Our BGNN can yield accurate and precise $\kappa$ posterior PDFs by processing all available photometric observations, e.g. the positions and magnitudes of individual galaxies around a line of sight. On average, it is 60\% more accurate than matching number counts $N_{\rm sim}$ and 30\% more accurate than matching distance-weighted number counts $N_{1/r, {\rm sim}}$. 
    \item When propagated into hierarchical inference, the BGNN-inferred posterior PDFs lend themselves to precise and accurate recovery of hyperparameters for a range of test populations shifted from the training distribution. On a population \liz{fully encompassed} by the training distribution $\kappa \sim \mathcal{N}(0, 0.02)$, the BGNN can recover the population mean on $\kappa$ precisely and accurately, translating into only a 0.1\% $1\sigma$ uncertainty contribution to $H_0$ without evidence of bias. The $N_{\rm sim}$ and $N_{1/r, {\rm sim}}$ summary statistics, on the other hand, underestimate $\kappa$, which would lead to a slight upward bias on $H_0$ even on this representative test set. 
    \item Both the BGNN and \chris{the simplified summary statistics} methods became more biased as the test populations sample the tails of the training distribution. The BGNN is found to be more robust to this training-test mismatch, however; on a particularly challenging shifted and narrow population with $\kappa \sim \mathcal{N}(0.04, 0.005)$, the bias on $H_0$ was 0.8\% for the BGNN as compared to 2\% and 1.5\% for the $N_{\rm sim}$ and $N_{1/r, {\rm sim}}$ summary statistics, respectively.
    \item Based on a simulated dataset consisting of 200,000 sightlines following $\mathcal{N}(0.01, 0.04)$, we have demonstrated the ability of BGNNs to effectively model the mapping between photometric observations and the continuous $\kappa$ space. On the other hand, \chris{the simplified} summary statistics matching \chris{as implemented in our paper} cannot extrapolate beyond the discrete $\kappa$ samples in the dataset, resulting in poor generalizability and accuracy. For both approaches, we expect the hierarchical inference performance to improve with bigger training sets that better sample high- and low- $\kappa$ environments. Further tests are needed to probe the relative performance of the BGNN and summary statistics methods with bigger training sets. 
\end{itemize}

\adriano{We have validated the BGNN method of $\kappa$ inference within the hierarchical Bayesian framework using simplified, well-controlled experiments. Being the first to infer the population $\kappa$ statistics, we took a pedagogical angle and focused on ``toy testbeds'' for our method. We assumed convenient Gaussian parameterizations throughout in order to conduct simple recovery tests, but it remains to probe the numerics involved with inferring non-Gaussian population $\kappa$ distributions, particularly those with a right skew. In addition, our test sets were drawn from the same simulation as the training set. In order to assess the impact of training-test mismatches in the assumed galaxy-halo connection, it will be important to test on sightlines derived from other simulations, such as the Millennium Simulation \citep{springel2005simulations,hilbert2009ray}. Ultimately, we can better control the associated bias by inferring extra hyperparameters governing various aspects of the galaxy-halo connection. These extensions and further validation tests are not limited to the BGNN method and apply for the summary statistics matching (ABC methods) as well.}


\acknowledgments
\section*{Acknowledgments}
This paper has undergone internal review in the LSST Dark Energy Science Collaboration. We would like to thank the internal reviewers Adriano Agnello, Elizabeth Buckley-Geer, and Christopher Fassnacht for their insightful comments. 

JWP developed the \textsc{Node to Joy} package implementing the data simulation, model training, and hierarchical inference; performed the analyses; and wrote the main text. 
SB contributed to the design, scope, analysis and writing. 
MU implemented the photometric noise model and performed systematics tests with sparse input features and varying photometric noise levels.
MC provided input on method development and model architecture.
AA advised on interpretation and presentation of results.
SWC provided input on hierarchical inference. SWC was supported by the KIPAC-Chabolla fellowship and NSF Award DGE-1656518.
PJM advised on project scope and design, and assisted in interpreting results and drawing conclusions.
AR advised on interpretation and presentation of results.






The DESC acknowledges ongoing support from the Institut National de Physique Nucl\'eaire et de Physique des Particules in France; the Science \& Technology Facilities Council in the United Kingdom; and the Department of Energy, the National Science Foundation, and the LSST Corporation in the United States.  DESC uses resources of the IN2P3 Computing Center (CC-IN2P3--Lyon/Villeurbanne - France) funded by the Centre National de la Recherche Scientifique; the National Energy Research Scientific Computing Center, a DOE Office of Science User Facility supported by the Office of Science of the U.S.\ Department of Energy under Contract No.\ DE-AC02-05CH11231; STFC DiRAC HPC Facilities, funded by UK BIS National E-infrastructure capital grants; and the UK particle physics grid, supported by the GridPP Collaboration.  This work was performed in part under DOE Contract DE-AC02-76SF00515.




\bibliography{main}

\section*{Appendix}

\section{Lensing formalism} \label{sec:lensing_formalism}
In this section, we review the gravitational lensing formalism for multiple lens planes, closely following the treatment in \cite{das2008large}. We begin with a general description in Section \ref{sec:general_lensing}, proceed to a widely used numerical approximation in Section \ref{sec:discretized_lensing}, and draw connections to our lensing quantities of interest in Section \ref{sec:lensing_quantities}. 

\subsection{General description} \label{sec:general_lensing}
Consider a photon that reaches the observer at an angular direction of $\bm{\theta}$ on the image plane. Suppose the source is located at radial comoving distance $\eta$ relative to the observer and at the angular position $\bm{\beta}$ on the source plane. The so-called lens equation maps $\bm{\theta}$ to $\bm{\beta}$:
\begin{align} \label{eq:lens_equation}
    \bm{\beta}(\bm{\theta}) &= \bm{\theta} - \bm{\alpha}(\bm{\theta}).
\end{align}
Here, $\bm{\alpha}(\bm{\theta})$ is the deflection angle due to lensing, which satisfies
\begin{align} \label{eq:deflection_lensing_potential}
     \bm{\alpha}(\bm{\theta}) = \nabla \psi(\bm{\theta}),
\end{align}
where $\psi(\bm{\theta})$ is the effective lensing potential out to comoving distance $\eta$. Defining $D(\eta)$ as the comoving angular diameter distance corresponding to $\eta$, we can express $\psi(\bm{\theta})$ as a two-dimensional projection of the Newtonian potential $\Psi$ \chris{(Equation \ref{eq:lensing_potential}}. Then Equation \ref{eq:lens_equation} can be rewritten as
\begin{align} \label{eq:lens_equation_integral}
    \bm{\beta}(\bm{\theta}) &= \bm{\theta} - 2 \int_{0}^{\eta} \frac{D(\eta - \eta')}{D(\eta') D(\eta)} \nabla_{\bm \theta} \Psi(\eta' \bm{\theta}, \eta').
\end{align}

\subsection{Multi-plane raytracing} \label{sec:discretized_lensing}
The integral in Equation \ref{eq:lens_equation_integral} is numerically solved by dividing the radial interval between the observer and the source into a finite number of concentric shells \citep{blandford1986fermat, schneider1992book, seitz1994some, jain2000ray}. Suppose there are $K$ such shells, indexed by $k$ in increasing order from the observer to the source. Denote the lower and upper comoving distances of shell $k$ as $\eta^{(k)}_L$ and $\eta^{(k)}_U$, respectively, such that $\eta^{(0)}_L = 0$ and $\eta^{(K-1)}_U = \eta$. The continuous deflection $\bm{\alpha}(\bm{\theta})$ experienced by the photon becomes approximated as discrete deflections $\bm{\alpha}^{(k)}(\bm{\theta}^{(k)})$. Equation \ref{eq:deflection_lensing_potential} turns into 
\begin{align} \label{eq:deflection_lensing_potential_discrete}
    \bm{\alpha}^{(k)}(\bm{\theta}^{(k)}) = \nabla \psi^{(k)}(\bm{\theta}^{(k)}),
\end{align}
where the discrete analog of the lensing potential $\psi(\bm \theta)$ in Equation \ref{eq:lensing_potential} is now 
\begin{align} \label{eq:lensing_potential_discrete}
 \psi^{(k)}(\bm{\theta}^{(k)}) = \frac{2}{D(\eta^{(k)})} \int_{\eta^{(k)} - \frac{\Delta \eta}{2}}^{\eta^{(k)} + \frac{\Delta \eta}{2}} d\eta' \Psi(\eta' \bm{\theta}, \eta').
\end{align}

Given the discretized deflection angles $\left\{ \bm{\alpha}^{(k)}(\bm{\theta}^{(k)}) \right\}$, we can raytrace the observed angular position $\bm{\theta}^{(0)}$ back through the series of lens planes:
\begin{align} \label{eq:raytracing_discrete}
    \bm{\theta}^{(k)} = \bm{\theta}^{(0)} - \sum_{i=0}^{k-1} \bm{\alpha}^{(i)}(\bm{\theta}^{(i)}),
\end{align}
for $k=1, \cdots, K$. In particular, for the source located at shell $k=K$, we have 
\begin{align} \label{eq:lens_equation_discrete}
    \bm{\beta} \equiv  \bm{\theta}^{(K)} = \bm{\theta}^{(0)} - \sum_{i=0}^{K-1} \bm{\alpha}^{(i)}(\bm{\theta}^{(i)}).
\end{align}
Naively applying the iterative scheme in Equation \ref{eq:lens_equation_discrete} is computationally prohibitive for large $K$. For deep, full-sky simulations, there exist iterative techniques that reduce the number of arithmetic operations and memory \citep{seitz1994some, jain2000ray, hilbert2009ray, hartlap2005studying,mccully2014new}. 

\subsection{Lensing quantities} \label{sec:lensing_quantities}
Our target lensing quantities are the convergence and shear, which describe the strength and direction, respectively, of the lensing effect along the entire line of sight. They can be identified in the distortion matrix $\Gamma$, obtained by differentiating Equation \ref{eq:lens_equation_integral} with respect to $\bm{\theta}$ \citep{schneider1992book}:
\begin{align} \label{eq:distortion_matrix}
    \Gamma_{ij}(\bm{\theta}) 
    &\equiv \frac{\partial \beta_i(\bm{\theta})}{\partial \theta_j } 
\end{align}

The lensing distortion can thus be decomposed into an isotropic change in area, parameterized by the convergence $\kappa=\kappa(\bm{\theta})$, and an area-preserving change in shape, parameterized by the shear components $\gamma_{1, 2} = \gamma_{1, 2}(\bm{\theta})$ forming the complex shear $\gamma = \gamma_1 + i\gamma_2$. In particular, $\psi$ satisfies the Poisson equation, i.e.
\begin{align} \label{eq:kappa_grad}
\kappa(\bm\theta) = \frac{1}{2} \nabla^2 \psi(\bm\theta).
\end{align}

Predicting the lensing signal of halos entails relating their mass density to $\kappa$. In the Born approximation limit, $\kappa$ is the weighted sum of the projected surface mass densities of individual halos. Denoting the three-dimensional mass density of the individual halos, indexed by $n$, as $\rho_n(\eta, \bm{\theta})$, we have
\begin{align} \label{eq:kappa_rho_integral}
    \kappa(\bm\theta) = \frac{4 \pi G}{c^2} \int_0^{\eta} d\eta' \frac{D(\eta -\eta')D(\eta')}{D(\eta)} \sum_n \int_{\mathbb{R}^2} d\bm \theta' \rho_n(\eta', \bm{\theta'}).
\end{align}

Note that the Born approximation is not strictly valid in the strong lensing regime, as the light paths undergo significant perturbations \citep{birrer2017line, 1996barkana}.

\subsection{Raytracing numerics} \label{sec:numerics}
Note that $\kappa$ and $\gamma_{1, 2}$ contain second-order gradients of the lensing potential $\psi$. In full-sky simulations, we can obtain these vector fields on the sphere by generating maps of the gradients with respect to the spherical harmonics. The computation can be performed rapidly using FFTs on sky pixelization schemes such as \textsc{HEALPix} \citep{hockney1988computer, lewis2005lensed}. The deflection field $\bm \alpha(\bm \theta)$ and distortion matrix $\Gamma(\bm \theta)$ are first evaluated on a mesh grid, after which multiple rays can be traced backward in parallel. The choice of mesh spacing, which we will call $\delta$, determines the spatial resolution of the projected matter density. 

In principle, all halos in the foreground of the source contribute to lensing, but the effect becomes smaller with increasing projected distance from the source. A cutoff in the multipole ($l_{\rm max}$) smoothes out the effects within these angular scales. The cutoff determines the ``field of view'' of the light cone, or the sky area around an angular position $\bm \theta'$ that contributes lensing potential to the calculation of $\Gamma(\bm{\theta'})$. Let us denote the average field of view as $D$. \textsc{CosmoDC2} uses $l_{\rm max}=8000$, which corresponds to $D \sim 1.5'$.

For smaller light cones with $D$ spanning a few arcminutes, flat-sky approximations may suffice. The field of view is divided into a rectilinear grid, from which the gradients are computed numerically using finite differences. Though not exact, an analog of $\delta$ in this approximation will be the grid spacing. Similarly, the size of the flat field is akin to $D$ operationally defined above.

\subsubsection{Structure-enhanced raytracing} \label{sec:structure_enhanced}
Defining the gravity-only component of our mock Universe is a large-scale cosmological N-body simulation called the Outer Rim \citep{heitmann2019outer}. It sampled a volume of (4.225 Gpc)$^3$ with 10,240$^3$ particles, yielding a mass resolution of $m_p = 2.6 \times 10^9 \rm{M}_\odot$. The \textsc{CosmoDC2} galaxy catalog was built on top of the Outer Rim particles using semi-analytic galaxy models \citep{skysim5000}. 

In addition to galaxy and halo information, \textsc{CosmoDC2} characterizes each galaxy with weak lensing shear and convergence. The lensing pipeline of \textsc{CosmoDC2} computed these quantities using a multi-plane raytracing algorithm outlined in Section \ref{sec:lensing_formalism}. In particular, the surface densities were evaluated on a \textsc{HEALPix} \citep{gorski2005healpix} grid of \texttt{NSIDE}=4096, which corresponds to a resolution of $\delta \sim 51''$ \citep{larseninprep}. 

For galaxy-scale strong lensing studies, we instead require lensing statistics that vary on smaller, $1''$ scales. At the same time, we must include the impact of large-scale structure at large angular scales, which are already captured in the \textsc{CosmoDC2} $\kappa, \gamma_{1,2}$ values. To obtain the structure-enhanced $\kappa_{\rm ext}$, we raytraced through the Outer Rim halos along the sightline of each source galaxy at a resolution of $\delta=1''$, while subtracting out the effect of any extra mass we were adding in the process. This subtraction scheme ensures that the mean curvature of the universe equals the one imposed by the background, which is true if and only if the mean convergence of all angular directions to all redshifts is zero \citep{birrer2017line}. That is, our structure-enhanced convergence values must satisfy
\begin{align} \label{eq:mean_curvature_assumption}
    \int d\eta \int d\bm{\theta} \kappa(\bm \theta; \eta) = 0,
\end{align}
where $\kappa(\bm \theta; \eta)$ denotes the convergence evaluated at the angular position $\bm \theta$ and comoving distance $\eta$. We assume that the \textsc{CosmoDC2} convergence values already satisfy this condition. In the following, we describe the process of structure-enhanced raytracing in detail.

Let us index each sightline by $s$. For a given source galaxy at angular position $\bm{\theta}_{s}$ and redshift $z_{s}$ defining the sightline, we first queried halos with redshifts $z < z_s$ and masses $M_{200}>10^{11} M_\odot$ located inside an aperture of fixed radius $1.5'$. These halos constituted the light cone. The aperture size was chosen to match the \textsc{CosmoDC2} field of view, $D \sim 1.5'$. The mass cut is a numerical choice, built in to reduce the computation time raytracing over halos with insignificant masses. We determined the mass threshold of $10^{11} M_\odot$ via a convergence test; it was increased from the base value of $10^{10} M_\odot$ until the $\kappa_{\rm ext}$ value calculated within the aperture remained constant. On average, there were about 500 halos per light cone. 

We then performed full multi-plane raytracing from the source plane $z_s$ through each of the deflector halo planes using the strong lensing simulation package \textsc{Lenstronomy} \citep{birrer2018lenstronomy}. Surface densities were evaluated at a resolution of 1$''$, representing a structure enhancement by a factor of $\sim51$ compared to \textsc{CosmoDC2}. We then evaluated the convergence at the center of our aperture, which we denote $\kappa_{\textrm{render}, s}$.

Adding $\kappa_{\textrm{render}, s}$ to the original \textsc{CosmoDC2} convergence $\kappa_{\textrm{DC2}, s}$ would capture both small- and large-scale fluctuations, as desired, but erroneously double count the Outer Rim particles already included in the $\kappa_{\textrm{DC2}, s}$ computation. To preserve the mean mass in the universe, we employ the following calibration scheme. We render the halos at random positions within the aperture, compute the resulting convergence at the aperture center for each realization $i$ which we denote $\kappa_s^{(i)}$, and compute the mean across $N$ realizations: 
\begin{align} \label{eq:mean_kappa}
    <\kappa_{s}^{(i)}>_i \equiv \frac{1}{N} \sum_{i=1}^N \kappa_s^{(i)}.
\end{align}
Moving the halos around corresponds to a Monte Carlo integration over all angular positions $\bm{\theta}$, i.e. \adriano{$\int d\bm{\theta} \approx \frac{1}{N}\sum_i^N$}. Our final, calibrated convergence value for a sightline $s$ results from subtracting off this quantity:
\begin{align} \label{eq:calibrated_kappa}
    \kappa_{s} = \kappa_{\textrm{DC2}, s} + \kappa_{\textrm{render}, s} - <\kappa_s^{(i)}>_i.
\end{align}

Assuming that we sample our sightlines uniformly across the sky, we have \adriano{$\sum_s \propto \int d\bm{\theta}$} and since we assumed that the \textsc{CosmoDC2} values satisfied Equation \ref{eq:mean_curvature_assumption}, we can claim
\begin{align} \label{eq:mean_curvature_requirement}
    \sum_s \kappa_{s} \approx \sum_s \kappa_{\textrm{DC2}, s} = 0. 
\end{align}
Combining Equation \ref{eq:calibrated_kappa} and \ref{eq:mean_curvature_requirement} gives an important validation test of our calibration scheme
\begin{align} \label{eq:test}
    \sum_s \kappa_{\textrm{render}, s} \approx \sum_s <\kappa_s^{(i)}>_i. 
\end{align}

\subsubsection{Halo rendering} \label{sec:halo_rendering}

To generate the halo catalogs, \cite{korytov2019cosmodc2} (henceforward ``\textsc{CosmoDC2} team'') ran a parallel, tree-based friends-of-friends (FOF) halo finder on the Outer Rim simulation with dimensionless linking length $b=0.168$ and minimum  requirement of 20 particles per halo. All particles, not just those in the original FOF halo, were counted in radial shells centered on the point of minimum potential. They then constructed halo merger trees using a particle-membership algorithm \citep{rangel2017building}. Their halo light cone was the product of tiling the simulation box in space to build a greater volume and applying a parallel solver that linearly interpolated halo positions between adjacent snapshot positions. The light cone filled one octant ($\sim \textrm{5,000 deg}^2$) of the sky and had a depth of $z = 3$.
Although the original \textsc{CosmoDC2} lensing pipeline evaluated surface densities directly from the Outer Rim particles, we take as input the auxiliary halo catalog generated from running a halo finder on the particles. We treat each halo in the catalog as a Navarro-Frenk-White (NFW) mass distribution \citep{navarro1997universal}. Its density can be written as 
\begin{align} \label{eq:nfw_param1}
\rho(r) = \frac{\delta_c \rho_c}{\left(\frac{r}{r_s}\right) \left(1 + \frac{r}{r_s} \right)^2}
\end{align}
where $\rho_c = 3H^2(z)/8 \pi G$ is the critical density, for the Hubble parameter $H(z)$ at the halo redshift $z$. The dimensionless characteristic density parameter $\delta_s$ sets the normalization. The scale radius $r_s$ is defined as the radius where the logarithmic profile slope $n_{\rm eff} = d \ln \rho /d \ln(r/r_s) = -2$. Note that the slope falls off as $n_{\rm eff} \rightarrow -3$ for $r/r_s \gg 1$. Here, $\delta_c, r_s$ are the two free parameters that characterize each halo.

An alternative, and arguably more interpretable, parameterization uses the halo concentration and the integrated halo mass, for some choice of the dimensionless overdensity parameter $\Delta$. The halo concentration $c_\Delta$ is a dimensionless shape parameter defined as
\begin{align} \label{eq:concentration}
c_\Delta \equiv \frac{r_{\Delta}}{r_s},
\end{align}
where $r_{\Delta}$ is the radius inside of which the halo mass density is $\Delta \times \rho_c$. Suppose we have the mass contained within $r_\Delta$, i.e. $M_\Delta \equiv \int_{0}^{r_\Delta} 4 \pi r^2 \rho(r) dr$. Given $M_\Delta$, the value of $r_\Delta$ can be determined using: 
\begin{align} \label{eq:nfw_rDelta_to_MDelta}
M_\Delta = \frac{4}{3}\pi r_\Delta^3 \rho_c \Delta \implies r_\Delta = \left(\frac{3}{4 \pi \rho_c \Delta} M_\Delta \right)^{1/3}
\end{align}
The halo concentration $c_\Delta$ is related to $\delta_c$ as
\begin{align} \label{eq:nfw_delta_c}
\delta_c = \frac{\Delta}{3} \frac{c_\Delta^3}{\ln(1+c_\Delta) - c_\Delta/(1+c_\Delta)}
\end{align}
\chris{The new parameterization is thus
\begin{align} \label{eq:nfw_param2}
& \rho(r) = \frac{\Delta \rho_c}{3} \left[ \ln(1+c_\Delta) - \frac{c_\Delta}{1+c_\Delta} \right]^{-1} \times  \left(\frac{r}{r_\Delta} \right)^{-1} \left( \frac{1}{c_\Delta} + \frac{r}{r_\Delta} \right)^{-2}
\end{align}}
Taking Equations \ref{eq:nfw_rDelta_to_MDelta} and \ref{eq:nfw_param2} together, the NFW profile with parameters $r_s, \delta_c$ in Equation \ref{eq:nfw_param1} can be described completely by $c_\Delta, M_\Delta$ instead. We make the conventional choice of $\Delta = 200$ in this paper. 

The \textsc{CosmoDC2} catalog lists the FoF mass with dimensionless linking length $b=0.168$ ($M_{\rm{fof}, 0.168}$) so we approximate $M_{200}$ with this mass value, since an exact conversion between $M_{\rm{fof}, 0.168}$ and $M_{200}$ does not exist. To assign the concentration $c_{200}$, we use the $c_{200} - M_{200}/M_\star$ fit derived from the Outer Rim halos \citep{child2018halo}:
\small
\begin{align}
    c_{200} &= A \left[ \left( \frac{M_{200}/M_\star}{b} \right)^m \left(1 + \left(\frac{M_{200}/M_\star}{b}\right)^{-m} \right) - 1\right] + c_0 \nonumber \\
    b &\equiv \frac{M_T}{M_{\star}}
\end{align}
\normalsize
where the fit parameters were $m=-0.10$, $A=3.44$, $b = 430.49$, and $c_0 = 3.19$. The individual dispersion in $c_{200}$ was $c_{200}/3$.

Granted, describing individual halos as spherical NFW profiles is an idealization. A more realistic description would be a prolate ellipsoid with a major axis length roughly twice as long as the minor axis \citep{jing2000density}. Halo shapes and profiles are also highly variable, depending in part on whether the halos are dynamically relaxed \citep{white2002mass, lukic2009structure}.

\section{Photometric catalog} \label{sec:galaxy_catalog}
\subsection{Galaxy-halo connection}
First, we briefly summarize how galaxies were painted on top of the halos in \textsc{CosmoDC2} to yield the galaxy catalog containing the broadband filter magnitude information. \liz{Readers are referred to \citet[\S5]{korytov2019cosmodc2} for further details. The galaxy-halo connection encoded in \textsc{CosmoDC2} is one that the BGNN, as well as the summary statistics matching, implicitly learns in order to map the photometric observation space to the $\kappa$ space.}

Producing the accompanying galaxy catalog from the halo catalog involved modeling the galaxy-halo connection with a hybrid of empirical and semi-analytic models. The UniverseMachine synthetic galaxy catalog provided the empirical model for predicting the star-formation history of galaxies \citep{behroozi2019universemachine}. It was chosen because it captures a wide range of statistics summarizing the observed galaxy distribution across redshift, including stellar mass functions, quenched fractions, and the dependence of two-point clustering to the star formation rate. Using the GalSampler technique, the \textsc{CosmoDC2} team matched every halo in the halo light cone to a suitable UniverseMachine \citep{behroozi2019universemachine} galaxy based on the halo mass ($M_{\rm halo}$) and assigned the galaxy's star-formation rate (SFR) and total stellar mass ($M_\star$) to the halo. By construction, this technique preserved the halo mass dependence of the two assigned properties, $P(\rm{SFR}, M_\star | M_{\rm halo})$, as well as that of the UniverseMachine halo occupation statistics, $p(N_{\rm gal}|M_{\rm halo})$. \adriano{For the UniverseMachine $M_\star - M_{\rm halo}$ relation for redshifts $z=0$ to $z=10$, see \S4.2 of \cite{behroozi2019universemachine}.}

To assign more complex properties, such as those due to galaxy mergers or metal production, the empirically-modeled galaxies were matched to those in the Galacticus library \citep{benson2012galacticus}. Each galaxy was modeled as a mixture of two Sersic components, i.e. a de Vaucouleurs bulge plus exponential disk. 
\subsection{Galaxy catalog properties}
The \textsc{CosmoDC2} galaxy catalog covered a 440 $\rm{deg}^2$ field spanning $0<z<3$. Each galaxy was characterized by various properties governing their surface brightness profiles. Of these properties, we took the projected coordinates (RA/dec) and LSST filter magnitudes ($ugrizY$) to construct the photometric catalog that was input to the network. More precisely, we drew sightlines at the positions of random source galaxies in \textsc{CosmoDC2} with fixed redshift $z_{\rm src} \approx 2$ and, for an aperture of fixed size around a given sightline, compiled the catalog information of galaxies within the aperture except the source galaxy. We included galaxies in the foreground of the source ($z < z_{\rm src}$), which contribute lensing, as well as the ones in the background ($z > z_{\rm src}$). The resulting sub-catalog represented the astrometry and broadband filter magnitudes of luminous tracers along a line of sight being queried. 

\begin{figure*}[!htb]
\includegraphics[width=\textwidth]{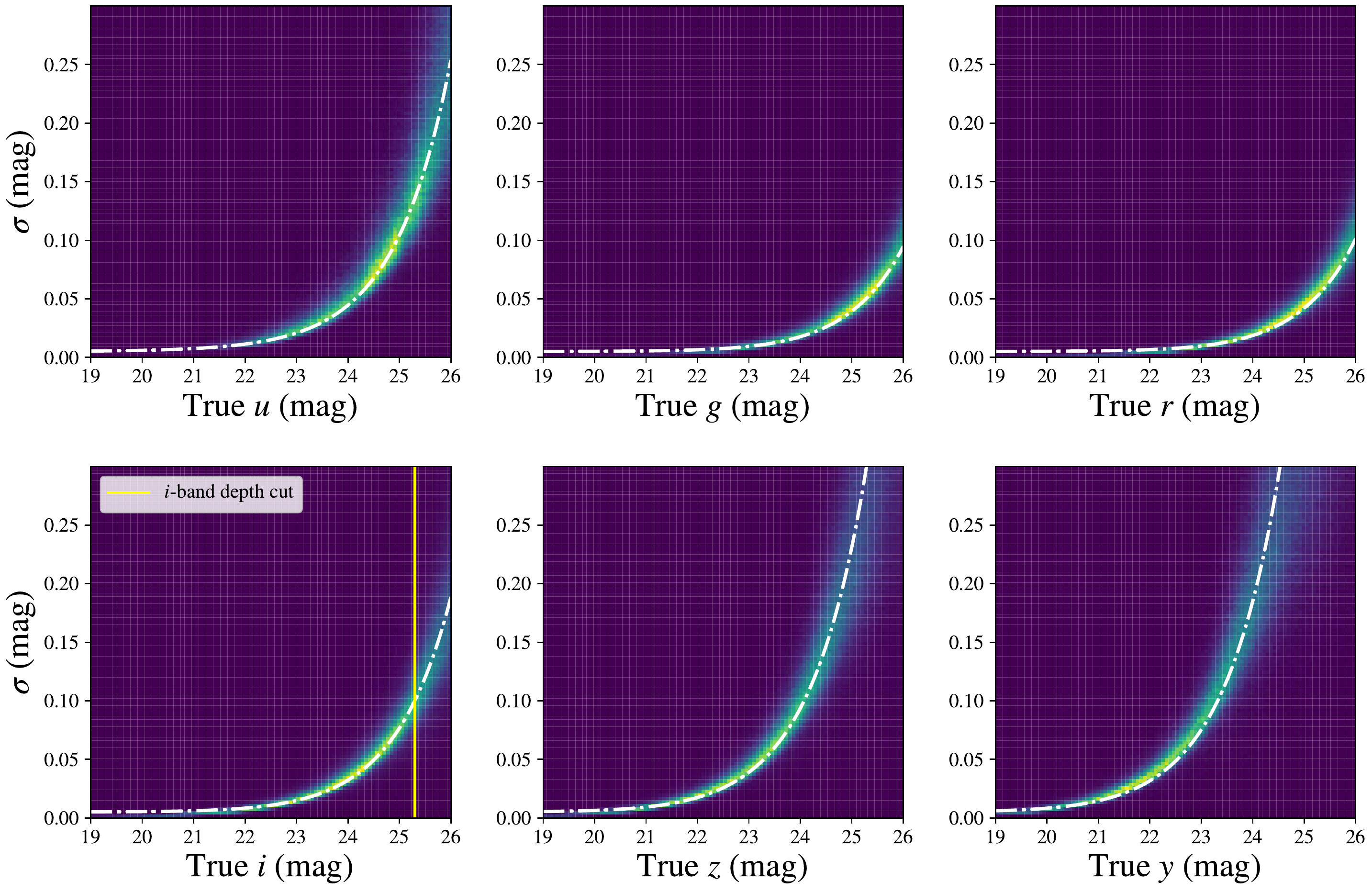}
\caption{We simulated Gaussian photometric errors, with 1-$\sigma$ uncertainties plotted in white dashed lines. Our error model agreed with the estimated uncertainty values in the DC2 catalog simulating 5 years of the LSST survey, shown in the heatmap. We also applied a cut of $i<25.3$ (yellow), corresponding to the $i$-band gold sample.}
\label{fig:gal_err}
\end{figure*}

\liz{To simulate photometric noise, we added magnitude-dependent LSST Y5 noise to the ground-truth $ugrizY$ magnitudes \citep{ivezic2019lsst}. The noise model is plotted in Figure \ref{fig:gal_err}. We applied a depth cut of $i < 25.3$, corresponding to the LSST $i$-band gold sample \citep{gorecki2014new}.}

\section{Experimental details} \label{sec:experiments}

\begin{figure*}[!htb]
\includegraphics[width=1\textwidth]{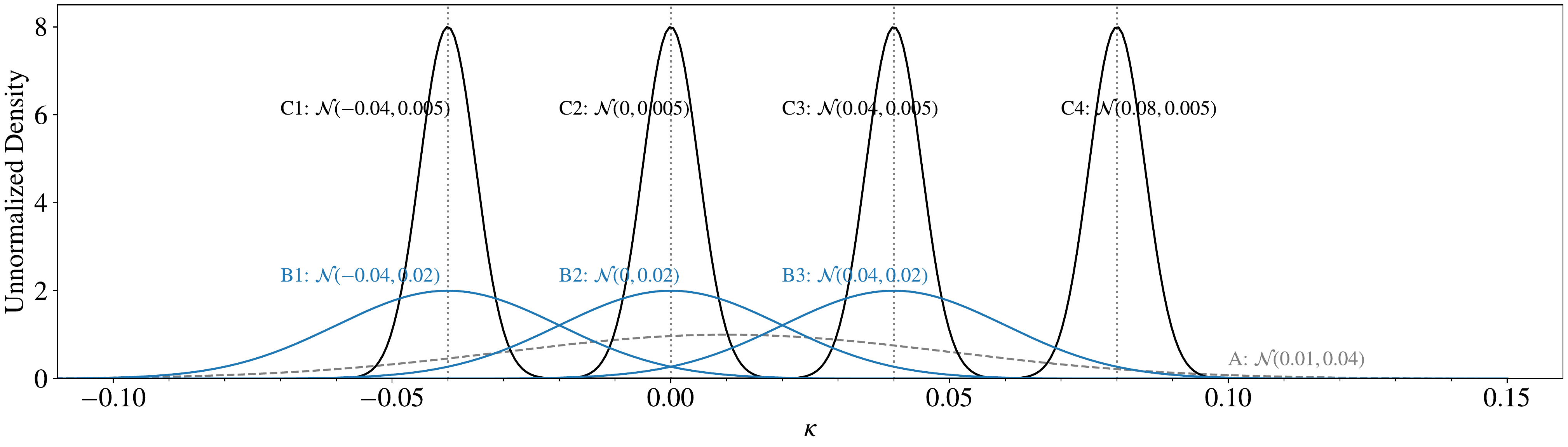}
\caption{Visualization of the test set distributions. A (dashed) refers to the training set distribution. The B group of experiments (blue, solid) probe varying $\mu_{\rm test}$ with broad $\sigma_{\rm test}$, and likewise with the C group (black, solid) with narrow $\sigma_{\rm test}$. Note that C4, in particular, has very little overlap with the training set.}
\label{fig:test_set_viz}
\end{figure*}

\liz{As a visual aid, we plot the training distribution in $\kappa$ (A) alongside the seven test distributions for B1-3 and C1-4 in Figure \ref{fig:test_set_viz}. Note that C4 has very little overlap with the training set.}

\end{document}